\documentclass[usenatbib]{mnras}

\usepackage[T1]{fontenc}
\usepackage{ae,aecompl}

\usepackage{graphicx}	
\usepackage{amsmath}	
\usepackage{amssymb}	
\usepackage{hyperref}  
\usepackage{times} 

\title[Oscillations  in   the light curves of 5 type 1 AGN]{Oscillatory patterns  in the light curves of 5 long-term monitored type 1 AGN}

\author[A. B. Kova{\v c}evi{\'c} et al.]{Andjelka B.  Kova{\v c}evi{\'c},$^{1}$\thanks{E-mail: andjelka@matf.bg,ac,rs}
Ernesto P\'erez-Hern\'andez,$^{2}$
Luka \v C.  Popovi{\'c},$^{1,3}$ 
\newauthor Alla I. Shapovalova,$^{4}$
 Wolfram Kollatschny$^{5}$
 and  Dragana   Ili{\'c}$^{1}$
\\
$^{1}$Department of Astronomy, Faculty of Mathematics, University of Belgrade, Studentski trg 16, Belgrade 11000, Serbia\\
$^{2}$Departamento de F\'{i}sica, Universidad de Guadalajara,  Revoluci\'{o}n 1500, 44420 Guadalajara, Jalisco, M\'{e}xico\\
$^{3}$Astronomical Observatory, Volgina 7, Belgrade 11060 , Serbia\\
$^{4}$Special Astrophysical Observatory of the Russian Academy of Science, Nizhnij Arkhyz, Karachaevo-Cherkesia 369167, Russia\\
$^{5}$Institut f{\"u}r Astrophysik, Universit{\"a}t G{\"o}ttingen, Friedrich-Hund Platz 1, 37077 G{\"o}ttingen, Germany}
\date{Accepted XXX. Received YYY; in original form ZZZ}

\pubyear{}

\begin{document}
\label{firstpage}
\pagerange{\pageref{firstpage}--\pageref{lastpage}}
\maketitle

\begin{abstract}

A new combined data of 5 well known type 1 AGN are probed with a novel hybrid method in a search for oscillatory behavior.
Additional analysis of artificial light curves obtained from the  coupled oscillatory models  gives confirmation for detected periods that could  have physical background.
We  find  periodic variations in  the long-term  light curves of 3C 390.3, NGC 4151,  NGC 5548 and E1821+643, with correlation coefficients larger than 0.6. We show that  oscillatory patterns of  two  binary black hole candidates NGC 5548 and E1821+643 corresponds to  qualitatively different dynamical regimes of chaos and stability, respectively. We demonstrate  that  absence of oscillatory patterns in Arp 102B  could be due to a  weak coupling between oscillatory mechanisms.  This is the first good evidence that 3C 390.3 and Arp 102B, categorized as double-peaked  Balmer line objects, have  qualitative different dynamics.
Our analysis  shows a novelty  in the oscillatory  dynamical  patterns  of  the light  curves  of these type 1 AGN.

\end{abstract}

\begin{keywords}
galaxies:active -- galaxies:nuclei --galaxies: Seyfert -- galaxies supermassive black holes --  methods: data analysis
\end{keywords}



\section{Introduction}

The optical emission of  Active Galactic Nuclei (AGN)  is of great complexity, containing  the stellar component, the emission lines from narrow line region and broad line region (BLR), the black body emission of  the accretion disc which is  orbiting a supermassive black hole (SMBH), the re-emission from the molecular torus  obscuring  the accretion disc, and synchrotron and inverse Compton emissions \citep[see e. g.][and references therein]{N13}. Hence, optical variability studies give insight in geometry,  physics and dynamics of  AGN   \citep{Re84, TC00, Haw02, Ses07, Mush11},  but the used data   must cover  long time periods \citep{Webb88}.  Because of this, AGN long-term monitoring programs  are crucial   for understanding physical aspects of  AGN as well as for establishing  cosmological constrains due to the wide distribution of AGN  over different cosmological time-scales \citep{Hon14, RiL17}.
Long-term  light curves are valuable to deal with open questions concerning AGN, particularly about possible periodicities and their physical origin \citep[see][and references therein]{Li16,Lu16}.
Searching for periodicity has been an important topic  of
AGN variability studies for about 4 decades, because confirmed
periodicity would strongly limit the possible physical
models and would help us determine the relevant physical
parameters in AGN \citep{L99}.
Additionally, oscillatory patterns immersed in AGN  light curves represent higher level structures in the  time series, which are suitable for comparative analysis of these objects.

The time series investigation methods  can be  classified as  shape-based, structure-based (or model-based), and dimensionality reduction \citep{Ding08}. 
The dimensionality reduction methods are  based on data transformation  such as discrete Fourier transformation, single value decomposition,  continuous wavelet  transformation (CWT) and discrete wavelet transformation (DWT), piecewise approximation, and Chebyshev polynomials.

Mostly, time-series analysis of AGN   attempts to characterize the similarity between curves   based  on their shape (i. e. utilizing Euclidean characteristics).
 For long time series, as  are   AGN long-term monitored data, structure-based and dimensionality reduction methods could be  more effective \citep{Ten00,Ke01,WQ08, Ag15}.
 Shape-based methods, being based on Euclidean metric, compare only time series of the same length, does not handle outliers or noise, and they are very sensitive to signal transformations,  phase shifts of signals, time delays, unsynchronized signals, etc. On the other hand, structure-based approaches  look for latent similarities, usually  by transforming  series into a new domain, 
where similarity can be more evident.
These  methods  can extract global features from the time series, and creating vectors of them can be also used to measure similarity and/or classify objects.  

A higher level information hidden in AGN light curves- oscillatory patterns    have been proposed to underlie  variability phenomena in numerous 
 AGN with the  extent  of  periods   from a few days to a few decades  \citep[see][and references therein]{Ch16}.
In other words, even if red noise is present in AGN light curves, there is a non-negligible  probability of detecting  periodic behavior \citep{Lei05}. 
On the other hand, different values for the periods can also be derived, which depends on the  data set and adopted  method  for testing \citep{KP06}. 
Thus it might be important to examine  whether the results remain robust  when the periodicity problem is investigated by a somewhat different method with limitations that light curves are combined from different monitoring campaigns.

Based on the above discussion, the aim of present paper is  twofold. Firstly, to utilize a new  data sets of  5 well known type 1 AGN,  combined from  several  well documented long term monitoring campaigns,  
 and then to  detect periodicity in the combined light curves using a novel method.
 Secondly, to construct models capable to reproduce oscillatory patterns  and phase space dynamics  of the combined light curves, and to find  physical relevance of  detected oscillatory patterns.

The paper is organized as follows. In section 2  we describe the used AGN sample, their  time series and method for detection of oscillatory patterns and phase trajectory calculation.
Section 3 focuses in detail on the results of the application of  proposed hybrid method for periodicity detection. Section 4 gives  a new comparative analysis of  oscillatory patterns and phase-space dynamics of combined and modeled light curves and  Section 5  summarizes the main conclusions.

\section{DATA  AND PERIODICITY ANALYSIS }

In the following, we briefly  describe data sets of  selected AGN, and then discuss methods and models  used for  the periodicity  analysis. 
In particular we present for the first time  the combined light curves: the continuum 6200 \AA\,,  and H$\alpha$ of  Arp 102B,  the continuum 5100 \AA \,,   H$\alpha$ and H$\beta$ of  3C 390.3 and  the continuum 5100 \AA\,,  and H$\beta$ of NGC 4151.
The combined light curves of NGC 5548 have been already published  as well as  the light curves of    E1821+643, so their plots will not be repeated here (see references listed inTable~\ref{tab:data}  for details).

\subsection{The  light curves}
An ideal data of ideal periodic phenomena  would perhaps satisfy conditions of  Nyquist theorem,
 so that  a light curve with the duration of  about  $1.5-2$ times longer than the  claimed periodicity  would be  enough for period measurement.
However, due to the  transient nature of underlying oscillatory patterns caused by  unfavorable characteristics of observed  light curves (e.g. signal to noise ratio, systematic errors, irregularities),  long term light curves are needed to establish a key aspect of periodic patterns in case of observed data \citep{Fan97}.
Existing optical light curves from different monitoring campaigns have been analyzed by different methods, but the information that can be derived from these observations has not been depleted.
One of  the possibilities is to combine these  data  into composite light curves, which  could benefit   longer time  baselines and 
better sampling.

The characteristics of these  objects  encompasses   the entire range of  type 1 AGN activity (see Table~\ref{tab:data}): from Seyfert 1 objects  NGC 4151 and NGC 5548 as prototypes of   spectral type variation (from Seyfert 1.0  to 1.8),
a very broad line and radio loud object  3C 390.3, then  Arp 102B that is considered as  a LINER and archetypical  disc accretion emitter due to extremely broad, double peaked Balmer lines, 
and finally  radio quiet but optically powerful near-by quasar E1821+643. 
Table ~\ref{tab:data}   lists general information for each object:
 object name, AGN type, redshift, total period of  monitoring programs, the combined light curve used for the analysis, mean sampling of the combined light curves,  as well as the variability parameter- excess variance  \citep[see eq. (7) in][and references therein]{Sim15}, and finally references from which the observations were taken.

 \begin{table*}
 
\centering
\begin{minipage}{\textwidth}
\caption{AGN sample for testing periodic variability. Columns are: object name,  AGN type,  redshift, total period of monitoring programs, the combined light curve, mean sampling of the combined  light curve, excess variance of the combined light curves, literature from which combined light curves were
compiled.}
\label{tab:data}
\begin{tabular}{cccccccc}
\hline
Object name & Type      & z        & Period    & CLC         & Sampling &   EV  & Reference\footnote{\noindent (1) \citet{Diet98}, (2) \citet{Sh10a}, (3) \citet{Diet12}, (4) \citet{Ser11}, (5) \citet{A15}, 
\noindent  (6)  \citet{W97}, (7)  \citet{OB98}, (8)  \citet{Sh13}, (9) \citet{Ser00}, (10) \citet{Kas96}, (11) \citet{Sh10b},
(12) \citet{Bon12}, (13)  \citet{Bon16}, (14) \cite{Sh16}.} \\
            &           &          &         &      &  {(days)}  &  &                                                      \\

        \hline
3C 390.3    & BLRG     & 0.056  & 1994-2014 & Continuum 5100 \AA& 11.6  & 0.1623 & 1, 2, 3, 4, 5 \\
            &           &          &         & H$\alpha$            & 34.5  & 0.1055 &                                                      \\
            &           &          &          & H$\beta$             & 20.5  & 0.1099 &                                                           \\
            &           &          & 1978-1996 & Continuum 1370 \AA & 64.4 & 0.1737 &6, 7                         \\
            &           &          &         & Ly$\alpha$           & 64.4  & 0.2539 &                                                                                                               \\
            &           &          &          & CIV                  & 64.4  & 0.2167 &                                                                                                           \\
Arp 102b    & LINER     & 0.024 & 1987-2010 & Continuum 6200 \AA & 78.1  & 0.0080 & 8, 9                                                                  \\
            &           &          &           & H$\alpha$            & 77.0  & 0.0245 &                                                                                                             \\
            &           &          &           & Continuum 5100 \AA & 73.0  & 0.0073 & 8                                                                                         \\
            &           &          &           & H$\beta$             & 60.0  & 0.0090 & 8                                                                                          \\
NGC 4151    & Seyfert 1 & 0.003 & 1993-2006 & Continuum 5100 \AA & 16.1  & 0.2847 & 10,11, 12                                                     \\
            &           &          &  1986-2006   & H$\alpha$            & 39.6  & 0.0740 &                                                                                                              \\
            &           &          &  1993-2006         & H$\beta$             & 19.2  & 0.1367 &                                                                                                            \\
NGC 5548    & Seyfert 1 & 0.017 & 1972-2015 & Continuum 5100 \AA & 6.9  & 0.0648 &13                                                                                            \\
            &           &          &           & H$\beta$             & 11.2  & 0.0917 &                                                                                                              \\
E1821+643   & Quasar    & 0.297 & 1990-2014 & Continuum 5100 \AA & 68.4& 0.0357 & 14                                                                                         \\
            &           &          &           & H$\beta$             & 68.4  & 0.0049 &                                                                                                              \\
            &           &          &           & Continuum 4200 \AA & 114.9 & 0.0359 &                                                                                                          \\
            &           &          &           & H$\gamma$            & 114.9 & 0.0356 &                                                                                                     \\ 
\hline
 \end{tabular}
 \end{minipage}

\end{table*}

\begin{figure}
	\includegraphics[width=0.5\textwidth]{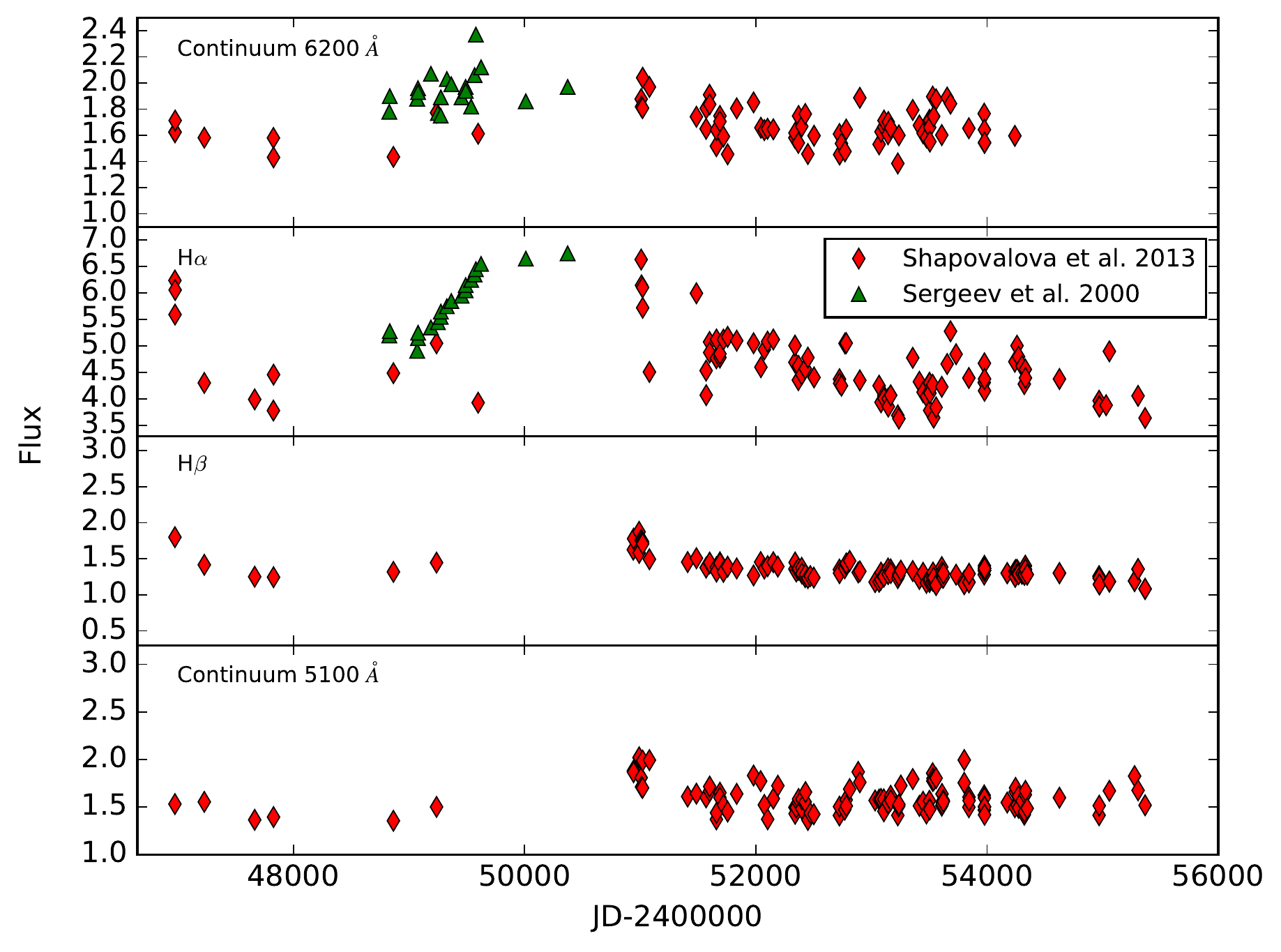}
    \caption{Combined  light curves  of  Arp 102B covering the period of 1987 - 2010. From top to bottom: continuum flux at 6200 \AA\,  (94 points),   H$\alpha$ and H$\beta$ line fluxes (110 and 141  points respectively), and continuum flux at  5100 \AA\, (116 points). The continuum fluxes are in units of $10^{-15} \mathrm{ergs\, s^{-1}\, cm^{-2}}$\AA$^{-1}$, and the line fluxes are in units of $10^{-13} \mathrm{ergs\, s^{-1}\, cm^{-2}}$. Observations from different campaigns are  marked  by different colors  given in legend on the second subplot from the top. }
    \label{fig:fig1}
\end{figure}

\begin{figure}
	\includegraphics[width=0.5\textwidth]{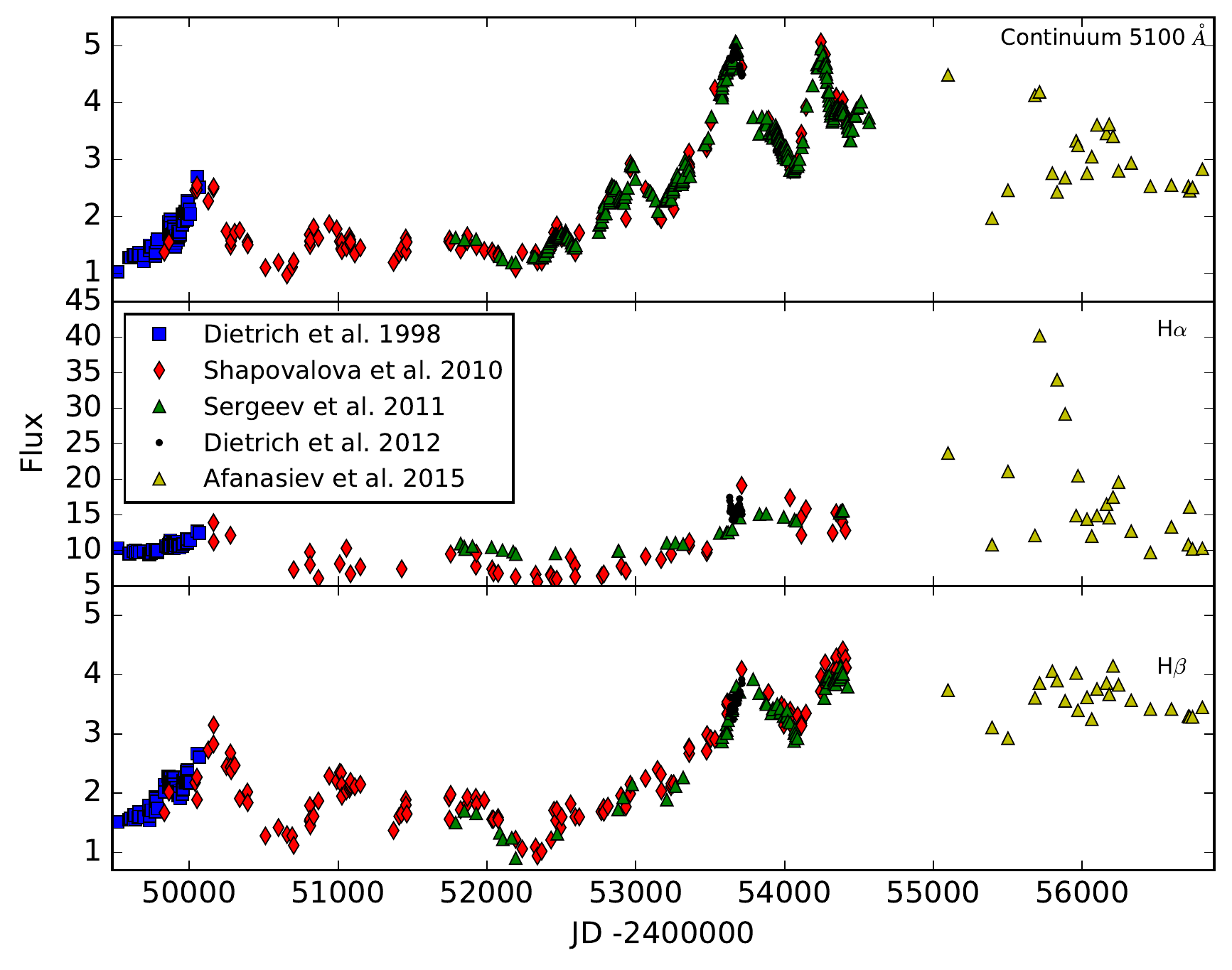}
    \caption{Combined  light curves  of  3C 390.3  covering the period of 1995 - 2014. From top to bottom: continuum  flux at 5100 \AA\,  (630  points),
    H$\alpha$  and H$\beta$ line fluxes (212 and 356 points, respectively).The continuum flux is in units of $10^{-15} \mathrm{ergs\, s^{-1}\, cm^{-2}}$\AA$^{-1}$, and the line fluxes are in units of $10^{-13} \mathrm{ergs\, s^{-1} \,cm^{-2}}$. Observations from different campaigns are  marked  by different colors  given in legend on the second subplot from the top. }
    \label{fig:fig2}
\end{figure}

\begin{figure}
	\includegraphics[width=0.5\textwidth]{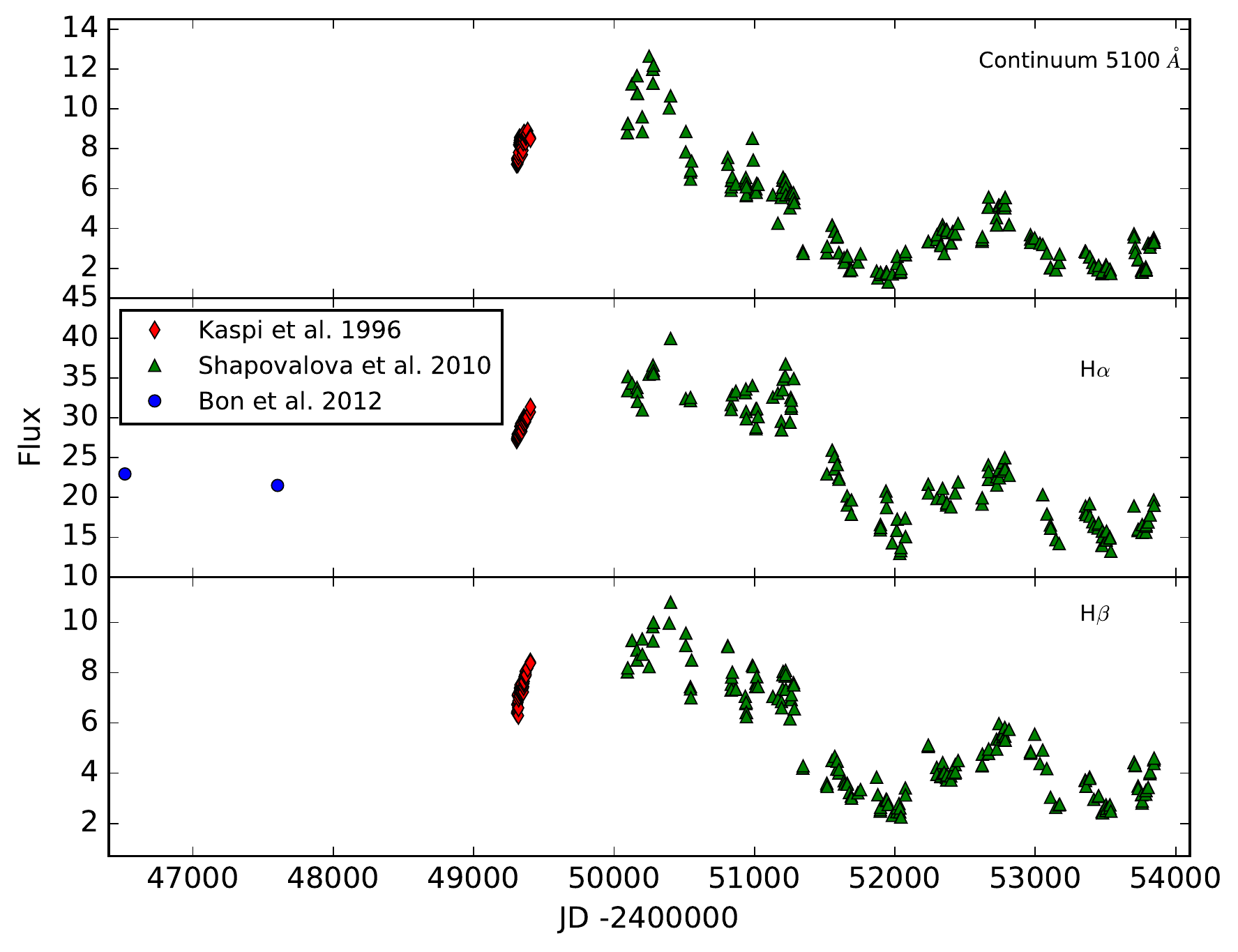}
    \caption{Combined   light curves  of  NGC 4151  covering the period of 1993 - 2010. From top to bottom: continuum flux  at  5100 \AA\,  (283  points),
    H$\alpha$  and H$\beta$ line fluxes  (168  and 238 points, respectively). The continuum flux is in units of $10^{-14} \mathrm{ergs\, s^{-1}\, cm^{-2}}$\AA$^{-1}$, and the line fluxes are in units of $10^{-12} \mathrm{ergs\, s^{-1}\, cm^{-2}}$.Observations from different campaigns are  marked  by different colors  given in legend on the second subplot from the top. }
    \label{fig:fig3}
\end{figure}

We note that the AGN Black Hole Mass Database \citep[AGNBHMD][]{BK15}  refers to a large number of published spectroscopic reverberation-mapping studies of  AGN.
But for our purpose  we combined  datasets for  each object as follows:
\begin{enumerate}
\item For   NGC 5548,  we used the  light curves   given  in  \citet{Bon16}. 
The   continuum 5100 \AA\,  and H$\beta$ line  are  covering  remarkable four decades long time span (see their Fig. 2);
\item For  Arp 102B (see  Fig.~ \ref{fig:fig1}), we  combined   datasets  of   \citet[][spanning the period of 1987-2010]{Sh13}  and  \citet[][covering the period of  1992-1996]{Ser00};
\item For 3C 390.3,  Fig.~ \ref{fig:fig2}, 
the  used observations are   from \citet[][covering period 1994-1995]{Diet98}, \citet[][covering period from September till December 2005]  {Diet12}, 
\citet[][covering period 1995-2007] {Sh10a}, \citet[][covering period 2000-2007]{Ser11}  and \citet[][covering period 2009-2014]{A15}. Due to complexity of this object, it might be useful
 to search for  periodicity information hidden  in the other wavebands, i. e. the  continuum 1370 \AA, Ly$\alpha$ and CIV lines collected  by  IUE satellite from 1978-1992  \citep{W97} and   from 1994 December to 1996 March \citep{OB98};
\item For NGC 4151,  Fig.~ \ref{fig:fig3},
the light curves  include  observations from \citet[][covering period over two months  in 1993]{Kas96}, from \citet[][covering period 1996-2006]{Sh10b}, 
and finally we added two observations  presented in \citet[][one from 1986, and other from 1989]{Bon12};
\item Up to now,  for E1821+643  the only information available are fluxes obtained from its the first long term (1990-2014) monitoring campaign conducted by  \cite{Sh16}, see their Fig. 5.
\end{enumerate}

 \begin{figure}
\includegraphics[width=0.5\textwidth]{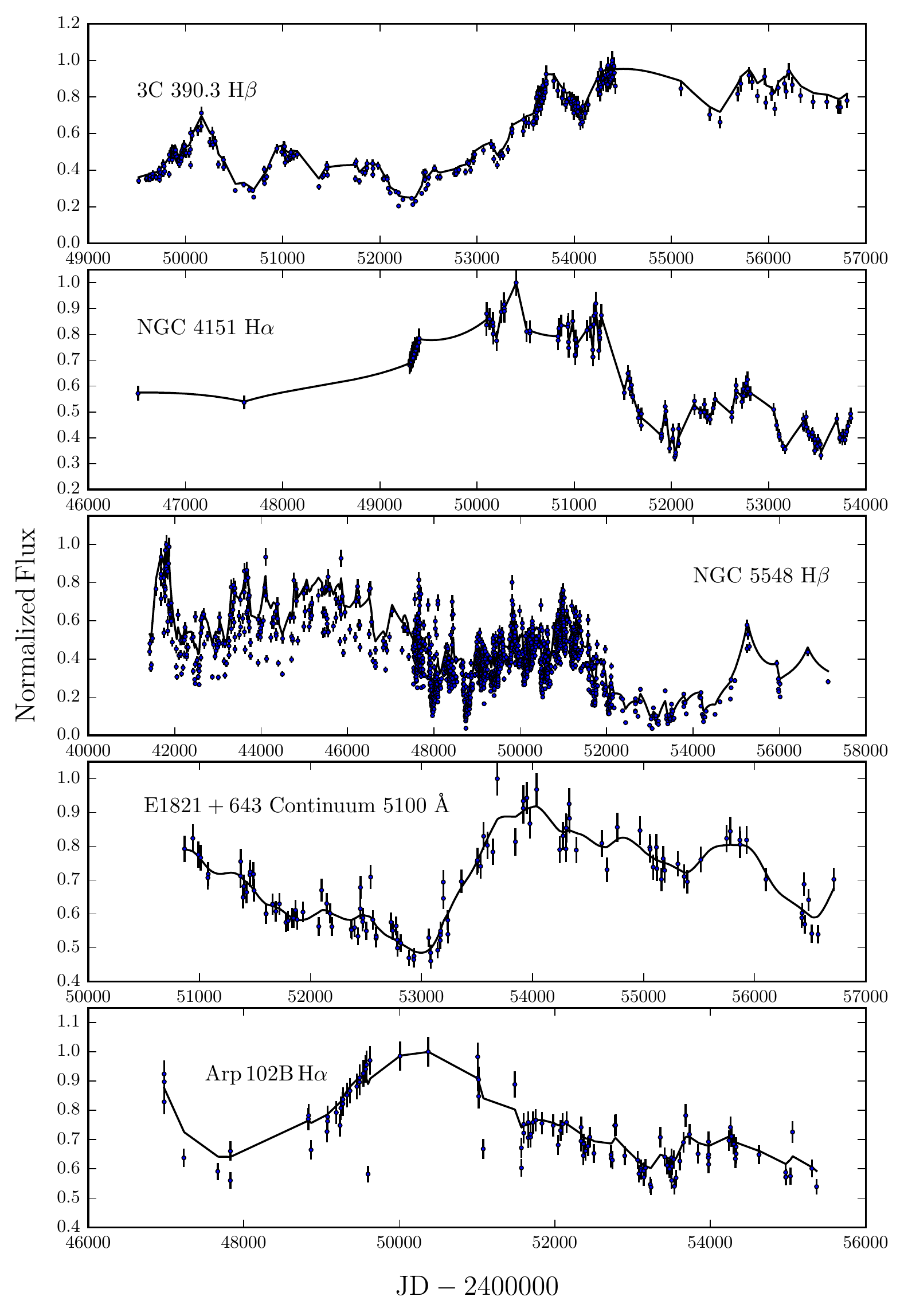}
\caption{ Examples of the comparison between the  GP posterior mean (solid line) and  observed light curves (dots) for the five objects. Names of the objects and light curves are given in each subplot.}
   \label{fig:sinusfit1}
\end{figure}

\subsection{Methods}

Almost all  AGN light curves  show neither apparent periodicity nor typical time-scale.
Their variations  look like each other  from largest (years) up to the smallest time-scales (hours).
Namely, AGN  fluxes exhibit self-similar or fractal variations \citep{KW99}.
On the contrary,  detection of a periodic variation in AGN light curves which would resemble the supermassive binary black hole  orbital time-scales are highly desirable \citep{Pop12}.
Such signals are weak, with long time-scale \citep[i. e. low frequency][]{N13} and immersed in the fractal and noisy structure of  AGN time series.
Usually,  low-frequency  filtering methods   require a certain prior knowledge of the expected range of periodic signal (such as frequency, phase, etc.).
 However,  problems we are encountering are  detection of a periodic signal in light curves when we do not know {\it a priori} that such signal exists or about  its characteristics.                            
\hfill\\
\indent To address such situation, we proposed a hybrid method to detect unknown weak periodic signals in the  AGN light curves, which is  a modification of the recently proposed timing methods used in geophysics \citep{Peri14}   and  acoustics research  \citep{YT05}.  \citet{Peri14} method identifies the fundamental frequencies of transient oscillations  in two time series by means of   Spearman's rank correlation coefficient analysis  of  
CWT of used data.
Their approach  is  a fusion of   the CWT  analysis (suited  for investigation of transient or unstable periodic phenomena), and Spearman correlation technique that accounts for non-linearity and variable amplitude of 
the wavelet coefficients.
\hfill\\
\indent  To account for the  irregular sampling in our data sets, we use  Gaussian process regression (GP), an approach that has been applied successfully in recent stochastic simulation/interpolation studies which  quantified  variability of quasar  light curves with arbitrary sampling \citep[see][and references therein]{Panc15a, Panc15b, Gri17}. Beside  the interpolation between data points, GP  also  
 makes self consistent estimates and allows us to incorporate the measurement errors while making minimal assumptions about the exact form of activity of the object.  It allows for a more flexible model which capitalizes on the data
 available and can model subtle but  important differences within the data.
To compensate the heavy influence of  the recent data which are better sampled, we interpolated  AGN light curves by means of GP obtaining  1-day  sampled    light curves.
This time interval is short enough in comparison to the long-term periods (years) and thus can barely  perturb the long-term variations.
The GP  approach  is based upon choice of a kernel.
For the calculations shown here, we utilized Ornstein-Uhlenbeck (OU) kernel, since sometimes it is a good descriptor of quasar variability \citep[see][and references therein]{Zh16}. 
 As a caution, we should note that  OU (or damped random walk) may not be ideal for Seyferts or  for all frequencies. For example \citet{Mush11} showed that the steep power spectra are inconsistent with the damped random walk, which 
is also confirmed  in the analyses by \cite{Koz17}  and \cite{Guo17}.
 For details about  OU covariance function (kernel)  see  \citet[][and references therein]{Kov17}.
   Moreover, as the light curves seem to be sparsely sampled over long monitoring  time, we show the examples of comparison between the GP  interpolated  and observed light curves  in  Fig. \ref{fig:sinusfit1}.  To exhibit parallels between fitted models,  the observed curves were normalized to maximal flux.
  \hfill\\
\indent  Instead of calculation of  CWT coefficients of the  time series,  our method determines  bandpass envelope of given curves  based on the  complex Morlet wavelet function. 
  \cite{YT05} have shown in acoustic analysis that this approach substantially mitigates overlapping and noise-induced signal.
We choose the complex mother wavelet function because it possesses two degrees of freedom (its real and imaginary part) to probe for structure in the  time series.
The complex Morlet wavelet transform of time series $ x(t) $ at an arbitrary scale $a$  and for translational parameter $b$ can be formulated as \citep{YT05}

\begin{equation}
\mathrm{CWT(a,b)}=\frac{1}{\sqrt{2\pi a}} \int^{+\infty}_{-\infty} x(t)e^{[(t/b)^2/a/2]\beta^2}e^{i\omega(t-b)/a}dt \\
\label{eq:eq1}
\end{equation}

\noindent where $i=\sqrt{-1}$,  $\omega$ is frequency of the wavelet function and $\beta$ is parameter controlling the wavelet function's shape.
Physically,
CWT(a, b) is the energy of $x(t)$ in scale $a$ at  time  $t = b$.
Then the envelope ($\mathrm{env(a,b)}$) of the wavelet coefficients are given by the  following metric expression \citep{YT05}

\begin{equation}
\mathrm{env(a,b)}=\sqrt{\mathrm{Re[(CWT(a,b))^{2}]+Im[(CWT(a,b))^{2}}]}\\
\label{eq:eq2}
\end{equation}

\noindent where $\mathrm{Re, Im}$ stand for  the  real and the imaginary part of a given CWT.  

The next step is to calculate correlation coefficients of the envelopes of the wavelet coefficients  of each light curve at each wavelet scale using Spearman rank correlation coefficient.
Spearman's coefficient  measure statistical dependence between two variables without a normality assumption for the underlying population (i.e. it is nonparametric measure).
Since it uses the ranks of the values in two variables, instead of their numerical values, it  is suitable for finding correlations in non-linear data and it is less sensitive to outliers.
A confidence interval for correlation coefficient  is constructed using Fisher's z transformation \citep[see][]{Holl99,Con99,Alt00}. The probability  p associated with Spearman correlation coefficient  is evaluated using an Edgeworth series approximation \citep[see][]{BR75}. 

The level  of match between oscillations presented in the two different light curves over a broad scale range can be visualized either as 3D correlation map (two axis are associated with periods in series and third  is corresponding correlation coefficients) or  in the form of 2D correlation map (two axes are corresponds to periods in two series while correlation coefficients are coded with different colors).
 Since  the values of the position parameter $b$ can be continuously varied,  and the scaling $a$ can be defined from the minimum (original signal scale) to a maximum chosen by the user, the CWT can be seen as a function of scales $a$ as it is shown in \citet{Grin04}. For the Morlet wavelet the period  is almost equal to the scale \citep[see][]{Grin04}. So the x and y axes of the correlation plots  depicts scales $a$, or equivalently, periods.
If the same period is presented in both light curves it will reveal itself in the high correlation regions centered on the 2D map's  diagonal.
The correlation between different periodicities (inter-oscillator correlation) would appear as regions of high correlation off the diagonal. The significance threshold for correlation coefficients  was set at 0.005.
We estimated periods by detecting peaks of correlation function which have largest correlation coefficients and p value bellow the significance threshold. The error of the resulting period ($\Delta$ P) was estimated  formally as the half-width of the corresponding peak \citep{Kud11}.

\subsection{Models}

One of  our ultimate  goals  is to   give a physical interpretation to the obtained results. To do so, 
we constructed  models that  are capable to produce oscillatory  and dynamical patterns  similar to those found in our  objects.

There are  two types of  such models: detailed and abstract \citep[see][and references therein]{Nak15}.
Detailed  models intend  to exactly reproduce as many characteristics of the observed system  as possible. 
Such models provide the quantitative understanding of the dynamical behavior of the studied  system. 
The other class,  abstract models   can capture some essential aspect of the system, such as rhythmic behavior. 
Their purpose is not  to faithfully  simulate all aspects of  dynamical behavior of  observed  system, 
but rather to describe  some universal aspect of its dynamics. 
Since it  is not focused on  detailed behavior of any specific system, but rather on the universal  characteristic of this behavior, 
it can give a unified frame for describing the behavior exhibited by a broad range of dynamical systems.
 Hence, such models allow us to accumulate a deeper  comprehension  of the general processes  existing in broad classes of systems.
For our purposes, the abstract  models  are suitable, simulating the  network of coupled oscillatory processes \citep{Pi01}. 
 In such a model, the evolution of each oscillatory system is described by three degree of freedom, the amplitude, period (frequency), and  the phase. 

So, to investigate whether interactions between oscillators  could contribute to the variety of oscillation patterns seen in AGN sample, 
we created two models.

 The first  type consists of two interacting units  $ U_{a}, U_{b}$, assuming that the interaction is linear and represented by the sum of one central  and one  remote oscillatory component.
The guiding equations are given as follows

\begin{equation}
\begin{aligned}
U_{a}(t)={}& A(t)\cdot \sin (2\pi f_{a} t +\phi)+cp_{b\rightarrow a}\cdot\\
                    & B(t)\cdot \sin(2\pi f_{b} t+2\pi f_{b}\tau) + W(t)\\
U_{b}(t)= {}&B(t)\cdot \sin (2\pi f_{b} t)+cp_{a\rightarrow b}\cdot \\
                    &A(t)\cdot \sin(2\pi f_{a} t+2\pi f_{a}\tau +\phi)+W(t)
\end{aligned}
\label{eq:eq3}
\end{equation}

\noindent where A(t)  and B(t)   are amplitudes of the  central and remote oscillatory process  before coupling occurs;  $U_{a}(t), U_{b}(t)$ are outputs of two  units at given time instance;
$f_{a},f_{b}$ denote the frequencies of interest in $ U_{a}, U_{b}$; $\phi$ is the phase difference;
$\tau$ is the delay between two units; $cp_{i\rightarrow j}$ is  the connection strength between $U_{i}$ to $U_{j}, i,j \in \left \{a,b\right \}$;  and $W(t)$ is  the red noise (i. e. Wiener process or Brownian motion). 
We generated W(t) on the time interval $[0, T]$ as a random variable depending continuously on all $t\in[0,T]$ and satisfying conditions:
 
\noindent $W(0)=0$, $W(t)-W(s)\sim \sqrt(t-s)N(0,1)$ for
$0\leq s<t\leq T$. 

\noindent $N(0,1)$ is the normal distribution with zero mean and unit variance, due to this fact, $W(t)$ is often called as Gaussian process. Note that for $0\leq s<t<u<v\leq T$, $W(t)-W(s)$ and $W(v)-W(u)$ are independent. For use in our model, we discretize $W(t)$ with time step $dt$ as $dW\sim \sqrt{dt} N(0,1)$ and found its cumulative sum. 

In difference to the first model, the second includes one central  and two remote   oscillatory component given as follows:
\begin{equation}
\begin{aligned}
U_{a}(t)={}& A(t)\cdot \sin (2\pi f_{a} t +\phi)+cp_{b\rightarrow a}\cdot\\
                    & B(t)\cdot \sin(2\pi f_{b} t+2\pi f_{b}\tau) +cp_{c\rightarrow a}\cdot\\
                                        & C(t)\cdot \sin(2\pi f_{c} t+2\pi f_{c}\tau_{1})+ W(t)\\
U_{c}(t)= {}&B(t)\cdot \sin (2\pi f_{b} t)+C(t)\cdot \sin (2\pi f_{c} t)+cp_{a\rightarrow b}\cdot \\
                    &A(t)\cdot \sin(2\pi f_{a} t+2\pi f_{a}\tau +\phi)+cp_{a\rightarrow c}\cdot\\
                    &A(t)\cdot \sin(2\pi f_{a} t+2\pi f_{a}\tau_{1} +\phi_{1}) + W(t)
\end{aligned}
\label{eq:eq4}
\end{equation}
\noindent Additional  remote oscillatory component has amplitude $C(t)$, frequency $fc$, coupling strength to central oscillatory process (and vice versa)  $cp_{a\rightarrow c}$, $ cp_{c\rightarrow a}$, phase $\phi_{1}$ and time delay $\tau_{1}$.
The values for the variables in both models were set to the following: all parameters were always  constant for considered time period, but were extracted from a normal Gaussian distribution for  100 reruns, separately  for  all involved oscillatory processes. In this way we have several degree of randomness: amplitudes, phases, red noise, and coupling parameter.
Each trial of  2000 time points was defined in both models with resolution of 1 arbitrary chosen time unit. 

To verify that similarities  between real and modeled correlations maps are not  accidental, 
 we determined how dynamics of   observed light curves  compare to the dynamics defined by  time-series models. 
To this intent, we calculated the  phase trajectories  of  observed and modeled data, since phases are  most sensitive to interaction, and provide description of connectivity within dynamical system which discloses a simple interpretation \citep{Kral11}.
The first step is to transform given  time series $y(t)=(y_{k} (t)), k=1,..N$ of each object into a cyclic observable.  This is completed  via construction of a two-dimensional embedding
 $(y, y^{H})$ \citep[see][and references therein]{Kral08}, where following equation 
 
 \begin{equation}
   y^{H}(t)=\frac{1}{\pi } PV \int_{-\infty}^{\infty}\frac{y(t)}{t-\tau}d \tau  
 \label{eq:eq5}
\end{equation}

 \noindent    defines  the Hilbert transform (HT)  of time series $y(t)$ and PV indicates the Cauchy principal value \citep{Bal09}.
  To better understand the meaning of the HT,  we can look at it  from the  point of view of a windowed sampling.
  The weight $\frac{1}{t-\tau}$ can be interpreted as a windowing width so  the HT is
 a convolution integral of $y(t)$ with $\frac{1}{t}$.  Therefore,  equation ~(\ref{eq:eq5}) emphasizes the local
properties of y(t).   
 To ensure existence of the integral, Lipshitz condition   $\left | y(t_{2})-y(t_{1})\right | \leq L \left |t_{2}-t_{1}\right| ^{\alpha}$  must be satisfied by $y(t)$ on given time interval $[t_{1},t_{2}]$,
 where L is a positive number. So if  $\alpha=0$ in an interval, $y(t)$ is discontinuous but bounded.
 We will not deal with Lipshitz condition here, however we will note that  the most optical  AGN time series satisfy  it, since L can be found as
a supremum of  all differences in  flux values. 
 
The original signal  ($y(t)$) and its
HT ($y^{H}(t)$) then formulate a complex analytic signal in the following form
 \begin{equation}
\tilde{y}(t)=y(t)+iy^{H}(t)
 \label{eq:eq6}
\end{equation}

\noindent where $i=\sqrt{-1}$.

\noindent  With this analytic signal, the time-dependent amplitude and phase
information embedded in the original signal can be easily extracted. Since the HT amplitude is an index of vibratory
energy and the phase is related to vibration frequency, such information is very desirable for characterizing dynamic
characteristics of a system.  HT is useful for long time series, low dimensional chaotic systems that exhibit transient chaos as well as for non stationary time series  \citep{LY03}.
Alternatively, one can use for $y^H$  the time derivative of y. 

The similarity of underlying dynamics in  observed  and simulated light curves can be
assessed by examining similarity of  their phase trajectories  in the phase space  $(y, y^{H})$.
 If the   oscillations are presented in the system, the phase  trajectory  in the plane  $(y, y^{H})$ will be closed  and can  fill a certain  annulus  of phase space, called system's attractor \citep{Nek15}.
 The absence of periodicity  will be manifested in phase space through a non closed phase curve. 
If  the underlying dynamics is  random, then the trajectory has no definite shape and spreads all over the space.
Moreover, if  phase trajectories  stay  within a finite (i.e., bounded) range of distance away from the critical point  they are  stable.
The phase space volume of  system with conservative dynamics is preserved as time evolves, while phase space of dissipative
dynamic system contracts as time evolves \citep{GOY87}.

 Finally, since determined periodicities are not obvious from visual inspection of observed light curves, it might be easier for the reader to discuss the reality of these variations on the basis of the comparison of the data and fitted sinusoids waveforms.
Therefore,  we applied multisinusoidal curves to perform the non-linear least square fitting of our data:
  \begin{equation}
        y={\sum\limits_{i=1}^{n} c_{i}\sin (\frac{2\pi t}{p_i}+\phi_i)+B}
         \label{eq:eq7}
\end{equation}
\noindent where $n$ stands for the number of detected periods in the light curves, $y$  denotes observed fluxes, and $t$ is corresponding time. From the fitting we estimate the following parameters:
 amplitudes $c_{i}$, periods $p_{i}$, phases $\phi_i$ and offset $B$ which is a parameter that handles measurement data with non-zero mean value.
 As the archetypal periodic function, sinusoidal signal is a good reference point for comparison to the observed light curves.
 The goodness of the  fit is assessed by estimation of $\chi^2$ and correlation coefficient between fitted and observed data.

\section{Results}
\begin{figure*}
\centering
\begin{tabular}{cc}
    \includegraphics[width=0.47\linewidth]{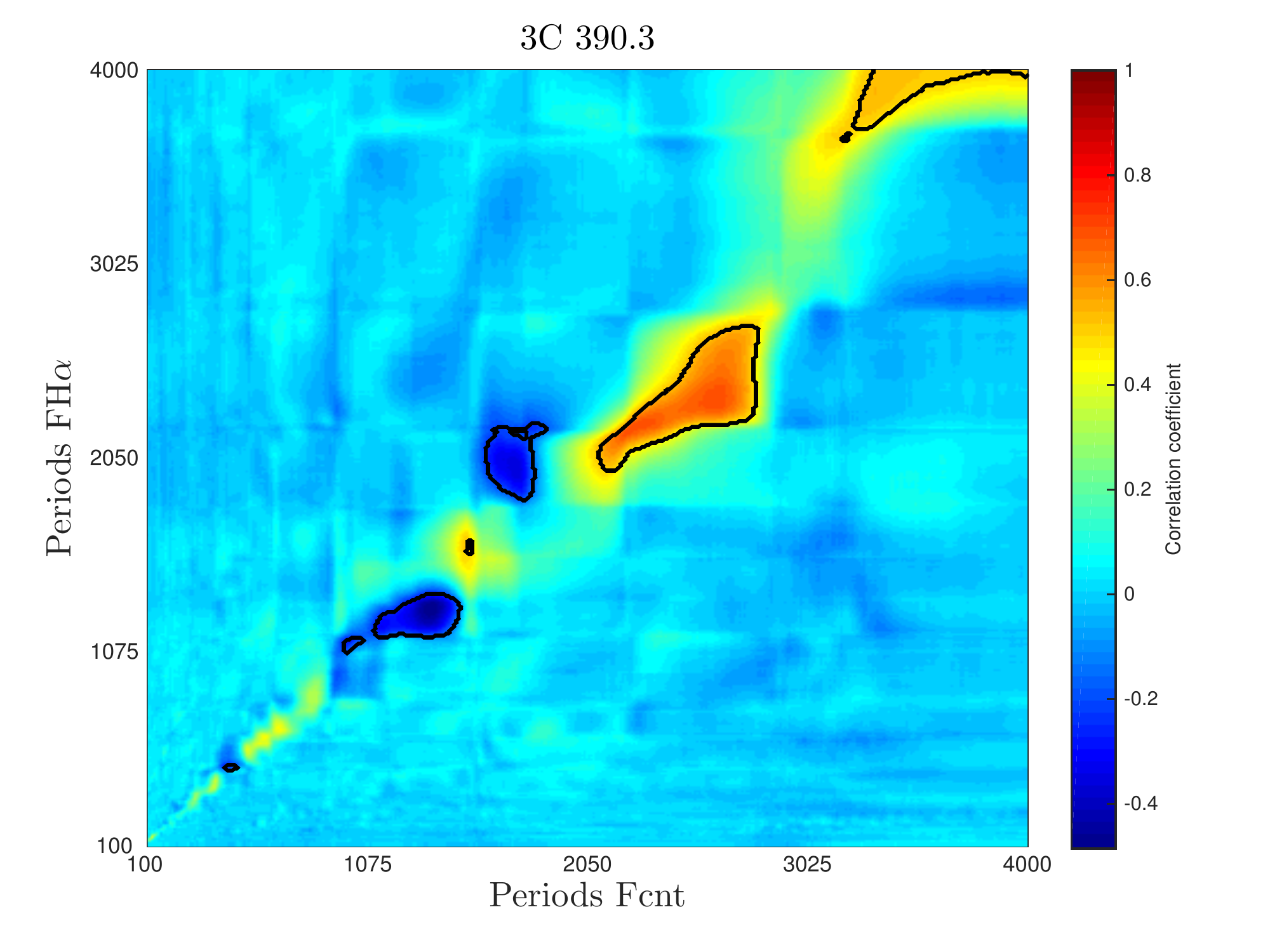}&
    \includegraphics[width=0.47\linewidth]{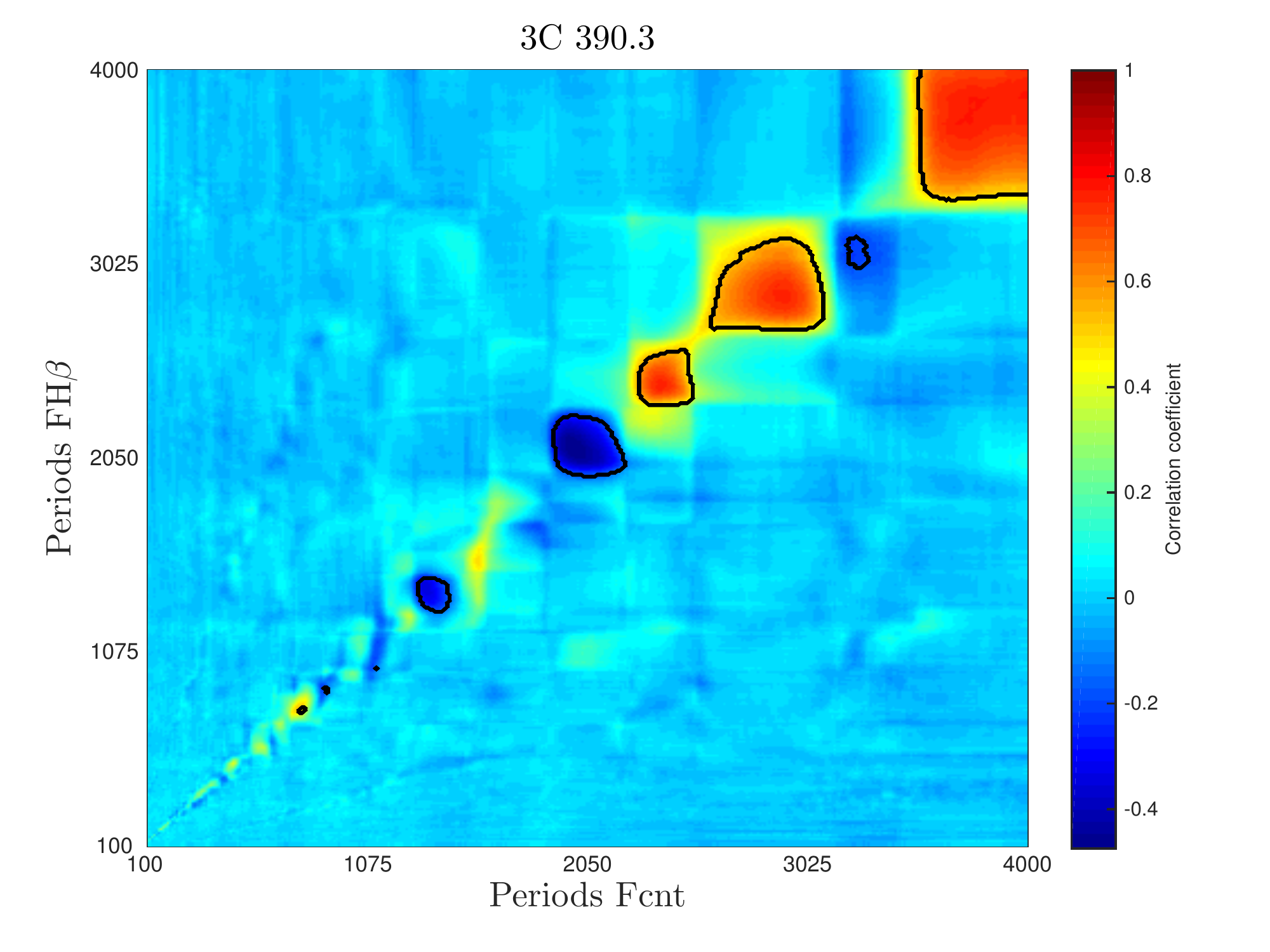}\\ [2\tabcolsep]
    \includegraphics[width=0.47\linewidth]{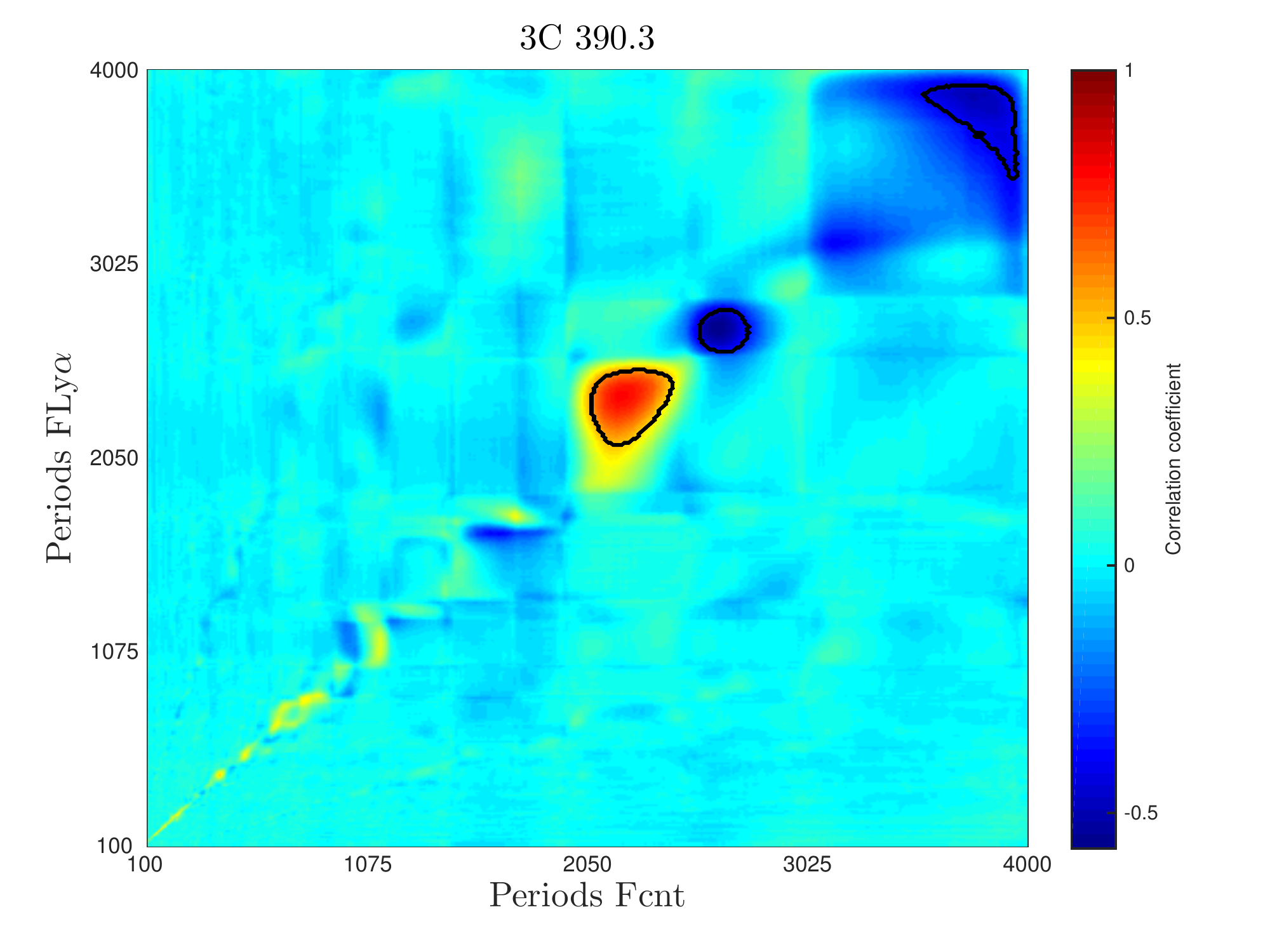}&
    \includegraphics[width=0.47\linewidth]{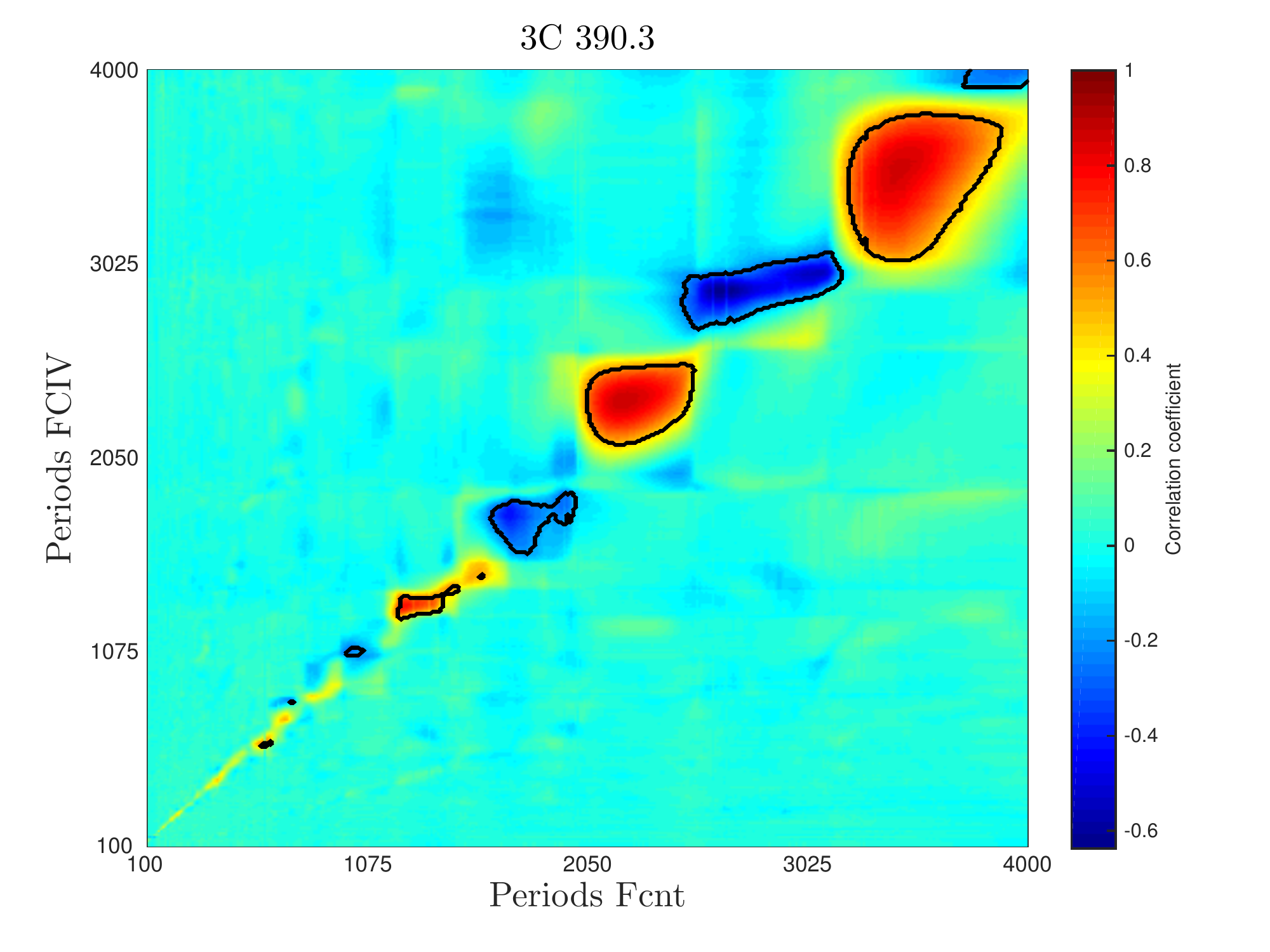}
\end{tabular}
\caption{2D correlation maps  of  all periodicities within the range $100-4000$ days found  in the light curves of 3C 390.3.   In   panels the horizontal axes are periods in the continuum  at 5100 \AA\,  and 1370 \AA, respectively.  Since correlation coefficients are symmetrical, the upper triangle of each  plot is a reflection  of  the corresponding lower triangle, however, it is presented for better visualization. 
The areas of high correlation are marked in  red. Note the prominent  blue  clusters of negative correlation.  Spurious non-physical signals (such as the dark region at about (1800, 3100) days, in the right bottom panel) appear as uncorrelated  (not reflected across the diagonal), because they are  uncoupled from real physical processes.
 Absence of off diagonal correlation clusters indicates that oscillations are caused by physical processes within 3C 390.3.}
  \label{fig:per3c}
\end{figure*}

\begin{figure}
    \includegraphics[width=\columnwidth]{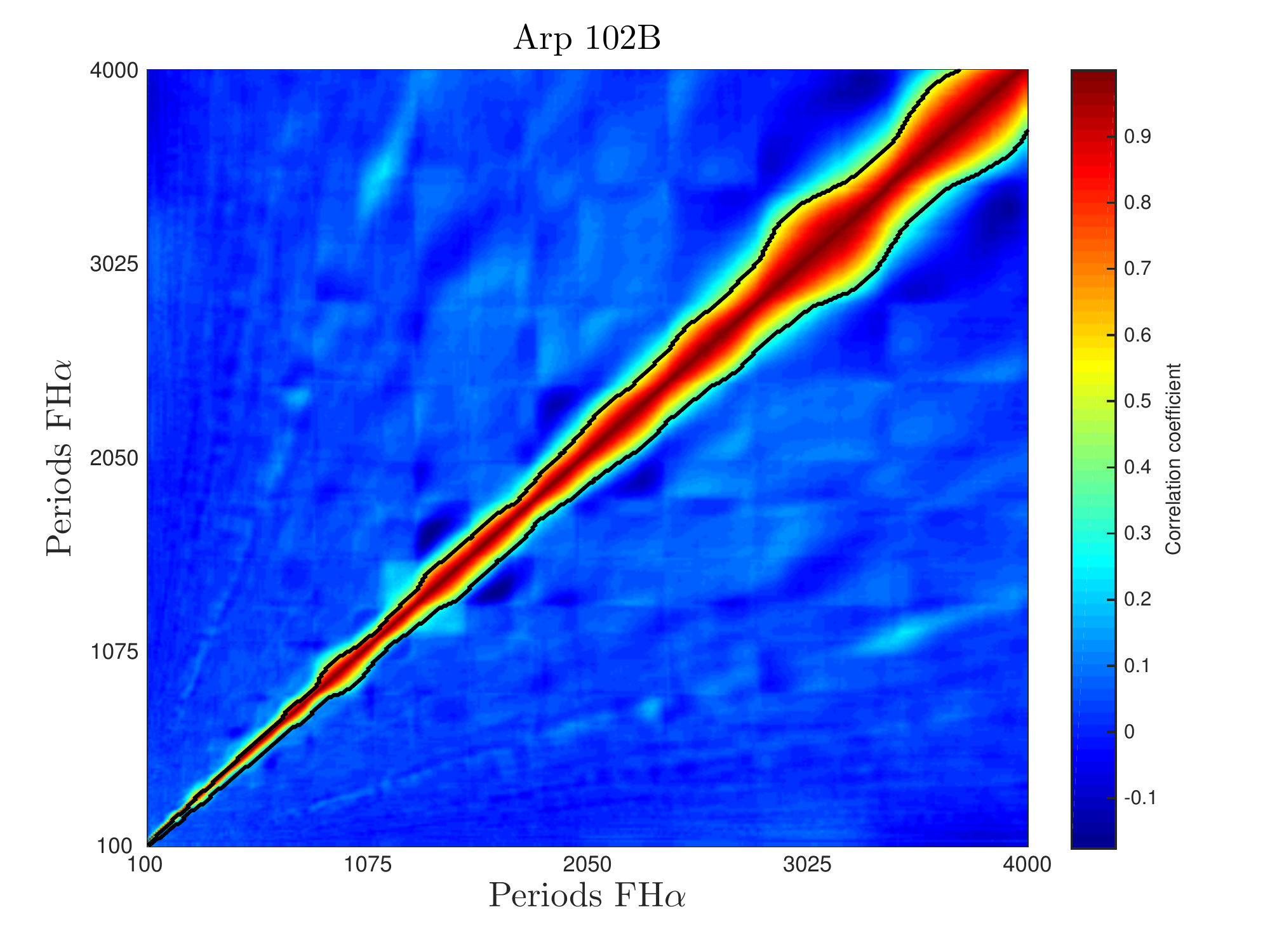}
\caption{As in Fig. (\ref{fig:per3c}) but for the   H$\alpha$  of  Arp 102B.  Note prominent stationarity of  diagonal correlation line and absence of correlation clusters.}
 \label{fig:perarp}
\end{figure}

Here, we provide  the main results of our periodicity analysis for each object   (Figs.~ \ref{fig:per3c}-\ref{fig:pere}).
First of all  we show the  novel  2D correlation maps
of   fundamental periodicities  for each object, which also  depict links between oscillations in the combined light curves. 
Then we  summarize all results in Table~\ref{tab:result}. 
 Since determined periodicities are not apparent  from the observed light curves, Fig.~ \ref{fig:sinusfit} compares them  with corresponding sinusoidal fit (Eq. ~(\ref{eq:eq7})). 
The estimated parameters of sinusoidal models are given in Table~\ref{tab:resultfit}.

 \begin{table*}
 
\centering

\caption{Periods in the combined light curves of our sample obtained with the hybrid method. Columns: object name, CLC1 and CLC2 are combined light curves used for periodicity analysis, $P\pm\Delta P$ is determined period and its formal error,
r is correlation coefficient corresponding to the period, $95\%$CI  is  $95\%$ confidence interval for r, p is significance i.e.  p-value  for r. }
\label{tab:result}
\begin{tabular}{ccccccc}
\hline
Object name & CLC1   & CLC2      &$ P \pm \Delta P$  & r  &$95\%$CI      & p                                               \\ 
   & && [yr]  &  &                                                      \\
\hline
3C 390.3    & Continuum  5100 \AA   & H$\alpha$    & $9.5 \pm0.3$& 0.5& (0.49,0.51) & $<0.00001$ \\
                              &         &           & $7.2\pm1.2$ & 0.69&(0.68,0.7) & $<0.00001$                                                 \\
                           &         &           & $6.3\pm0.9$ &0.68& (0.67,0.69) & $ <0.00001$                                                   \\
             &         &           & $4.0\pm0.04$ &-0.47& (-0.48,-0.45) & $ <0.00001$                                                   \\
             &         &           & $5.44\pm0.1$ &-0.35& (-0.37,-0.33) &$ <0.00001$                                                   \\
\cline{3-7}
                 &          & H$\beta$             & $10.11\pm0.1$  & 0.77 & (0.76,0.78) &   $<0.00001$                                                         \\
                               &          &                    & $7.67\pm0.02$  & 0.71 & (0.7,0.72)  &$<0.00001$                                                         \\
                            &          &                    & $6.42\pm1.6$  & 0.75 & (0.74,0.76) & $<0.00001$                                                         \\
                              &          &                    & $5.43\pm0.8$  & -0.47&  (-0.48,-0.45)& $<0.00001$                                                         \\
                         &          &                    & $3.6\pm0.4$  & -0.33&  (-0.35,-0.31) &$<0.00001$                                                         \\
\cline{2-7}
                & Continuum 1370 \AA  &Ly$\alpha$    & $10.34\pm0.1$ & -0.47& (-0.49,-0.45)   &$<0.00001$                        \\
              &         &       &$ 7.1\pm0.02$  & -0.53 & (-0.54,-0.51)  &$<0.00001$                                                                                                             \\
                            &         &       & $6.25\pm1.42$  & 0.77 & (0.76,0.78)  & $<0.00001$                                                                                                             \\
\cline{3-7}
               &          & CIV                  & $9.42\pm0.02$ & 0.85 &  (0.84,0.86)  &  $<0.00001$                                                                                                      \\
      
              &          &               & $7.84\pm0.02$& -0.6 &  (-0.61,-0.59)  &  $<0.00001$                                                                                                      \\
                  &          &               &$ 6.4\pm1.22$& 0.85 & (0.84,0.86)  &     $<0.00001$                                                                                                      \\
          &          &               & $4.68\pm0.7$& -0.42 &  (-0.44,-0.40)    &$<0.00001$                                                                                                      \\
              &          &          &   $ 3.4\pm0.4$&    0.75   &(0.74,0.76)      & $<0.00001$                                                                                                  \\
\hline
\hline
Arp 102B    &  Continuum 6200 \AA   &  H$\alpha$  &-& -    &   -                                                           \\
            &       Continuum 5100 \AA     &      H$\beta$      &          - & -  &-                                                                                      \\
\hline
\hline
NGC 4151    &  Continuum 5100 \AA &H$\alpha$  & $13.76\pm3.73$  & 0.96 &(0.956,0.962)  & $<0.00001$                                                     \\
       & && $8.33\pm2.33$ & 0.97 &(0.968,0.972)    &$<0.00001$                                                     \\
    & && $5.44\pm1.29$ & 0.98 &  (0.978,0.981)  & $<0.00001$                                                     \\

\hline
 \hline  
   
     NGC 5548    &Continuum 5100 \AA  & H$\beta$   & $13.3\pm2.26$ & 0.87& (0.867,0.873)  &$<0.00001$                                                                                            \\

  \hline
  \hline

           E1821+643   & Continuum 5100 \AA   &H$\beta$  & $12.76\pm5.6$ & 0.98 & (0.979,0.981)  & $<0.00001$                                                                                     \\
 &   &  & $6.93\pm 1.99$ & 0.80 &  (0.792,0.808)  & $<0.00001$                                                                                     \\
        &   &  &$ 4.75\pm0.79$ & 0.80& (0.792,0.808)  & $<0.00001$                                                                                     \\

  \cline{2-7}     
                    &  Continuum 4200 \AA         &  H$\gamma$        &     $12.36\pm6.1$      & 0.99 & (0.989,0.991) & $<0.00001$                                                                                                            \\
                                 &          &          &     $6.52\pm3.26$     & 0.91 &(0.906,0.914)  & $<0.00001$                                                                                                            \\

                     &          &          &     $4.34\pm0.74$      & 0.94 & (0.937,0.943)  & $<0.00001$                                                                                                            \\

\hline
 \end{tabular}
\end{table*}

 \begin{table*}
 
\centering

\caption{The estimated parameters of  sinusoidal best-fitting  (Eq. ~(\ref{eq:eq7})) of  normalized observed light curves. Columns: object name, light curve,  amplitudes $c_i$ , periods $p_i$, phases $\phi_i$, offsets $B$, the correlation coefficient between the modeled and observed  light curves r, and   chi-square goodness of fit $\chi^2$. Each line represents a set of  parameters for one sinusoid.} 
\label{tab:resultfit}
\begin{tabular}{cccccccc}
\hline
Object name & LC   & $c_i$      & $p_i$  & $\phi_i$  & $B$    & r    &$\chi^{2}$                                           \\ 
   & && [days]  & [radians] &        &   &                                           \\
\hline
3C 390.3    &H$\beta$  & $0.11 \pm 0.02$  & $3760 \pm 7$ &$6.02  \pm 0.01$ & $0.52\pm 0.01$&0.81 & 4.748 \\
                      &         &  $0.05\pm 0.03$         & $2743\pm 15$ &$ 5.51  \pm 0.03$& &                                                \\
                       &         & $ 0.29 \pm 0.04$           & $2300\pm 2$  &$5.47  \pm 0.03$& &                                             \\
                       &         & $0.17 \pm 0.03$           &$ 2000\pm 2$ &$0.17  \pm 0.005$ & &                                                   \\
                       &         &     $ 0.08 \pm 0.01$       &$1322\pm 1$ &$-5.24 \pm 0.1$ &  &                                                   \\

\hline
\hline
NGC 4151&   H$\alpha$  & $0.22\pm0.01$& $5580\pm 435$   &$1.52  \pm 4.34$ &0.63 $\pm$0.02  &  0.96&    0.381                                           \\
       & &$-0.07\pm 0.02$  & $2730\pm 422$ &$-4.20  \pm 5.63$ &   & &                                              \\
    & & $ -0.08\pm 0.01$ &$1534\pm28$  &$-4.02  \pm 3.82$&    & &                                                    \\
\cline{3-8}
&   H$\alpha$  & $-0.23\pm0.01$& $5165\pm3$   &- &0.63 $\pm$0.01 &  0.87&    1.275                                           \\
\hline
 \hline  
   
     NGC 5548    & H$\beta$   & $ -0.10\pm 0.01$ & $4378 \pm70$ &-5.35$\pm$1.12 &0.40$\pm$0.004&0.40&32.804\\

  \hline
  \hline

           E1821+643   & Continuum    &  $0.16\pm 0.001$& $4511\pm 1$   &$0.02  \pm 0.005$ &0.71$\pm$0.0  &  0.87&   0.449                                           \\
        &5100 \AA &$0.50\pm 0.0002$  & $2529 \pm 0.005$ &$ 1.57 \pm 0.03$ &   & &                                              \\
    & & $ 0.07\pm 0.005$ &$1977\pm0.1$  &$ 1.10 \pm 0.002$&    & &                                                    \\

\hline
 \end{tabular}
\end{table*}

\indent  \textbf{3C 390.3.} 
 The  2D correlation maps in Fig.~ \ref{fig:per3c} show the signature 
 of the   6.3  and 7.1 yr period to be  present in  the  continuum and H$\alpha$  line. 
  The method yields $\sim 0.7$ for the correlation coefficients of both periods with a significance $p<0.00001$.
 Somewhat weaker periodicity  (correlation coefficient  $\sim$ 0.5, but still significant)  was found  at  9.5 years.
 Note, however, that on the same significance level the periods of  4  years and 5.4 years are seen in two light curves but with negative correlation coefficients.  
 Similarly,  in the  continuum and H$\beta$  line periods of 10, 8 and 6.4 years  are identified with even larger  correlation coefficients  than in previous case, while periods of 5.4 and 3.6   expressed  negative correlation coefficients as in the case of the continuum and H$\alpha$ line.  
 These 'islands' of negative correlation are totally surprising  and worthy of further investigation. 
 We found significant periods of    9.4, 6.4, and  3.4 yr (correlation coefficients around 0.8)  in the continuum  1370 \AA\,and $CIV$, while negative correlation, stronger then in the case of  H$\alpha$ and H$\beta$ emission lines,  appears at 7.8 and 4.7 yr.
  On the other hand, largest negative links between oscillations in the continuum  1370 \AA\,and Ly$\alpha$ are seen at 7.1 yr and 10.3 yr,  while positive relationship is seen at 6.3 yr.
 The periods  are  consistent across all light curves except that correlation coefficients changed polarity differently  in the case of IUE light curves.
   A visual inspection of Fig. \ref{fig:sinusfit} shows that multisinusoidal models are quite successful in describing the peaks and troughs of the original data.
 This is reflected in high correlation between the fitted model and original data (see Table~\ref{tab:resultfit}). However, $\chi^2$ is large due to inaccurate reconstruction in the large gapped period, because  of the lack of data.

\indent \textbf{Arp 102B.} While every period  is incidentally  correlated with itself (correlation coefficient  equals  1 on the diagonal), any transient oscillations having physical origin which are  hidden within the light curves should be reveled  as clusters close to the diagonal. The main common feature in all correlation maps  is the absence of any  periodic variability (i. e correlation  clusters), even in autocorrelation maps. This is well illustrated in   Fig.~ \ref{fig:perarp}.

\indent \textbf{NGC 4151.} 
The data set  of    the H$\alpha$ line is  long enough, for  our hybrid method to clearly detect periods  at significance level $p<0.00001$.
Obtained autocorrelation map  (see Fig. \ref{fig:per41}) revealed three clusters of significant periods
 (large correlation coefficients)  at 5.4, 8.3 and 14 yr in cadence.
  Phase curves of  the continuum and H$\beta$ line are quite similar to phase curve  of the  H$\alpha$ line but their main loops are open. Due to this discrepancy in  dynamics, detected periodicities of  $\sim$8 yr
 and $\sim$5 yr (corresponding to the smaller loops in phase curves) in the continuum and H$\beta$ line are not reported in Table~\ref{tab:result}. H${\alpha}$ is only longer than the continuum and H${\beta}$ due to the
 two data points (MJD 46511.49 and 47603.29) separated  by $\sim 3$ years. Exclusion of these data points from  the  H${\alpha}$ line produces open phase curve.  It appears that these two points have quite an effect on
 the topology of  phase curves.
 { The associated multisinusoidal fit  is estimate as in the earlier case. The three sinusoid model is undoubtedly  superior to one sinusoid model  as is evident both, from the correlations between fitted and observed values and from the $\chi^2$ parameter (see Table~\ref{tab:resultfit} and   Fig. (\ref{fig:sinusfit})) .

\begin{figure}
    \includegraphics[width=\columnwidth]{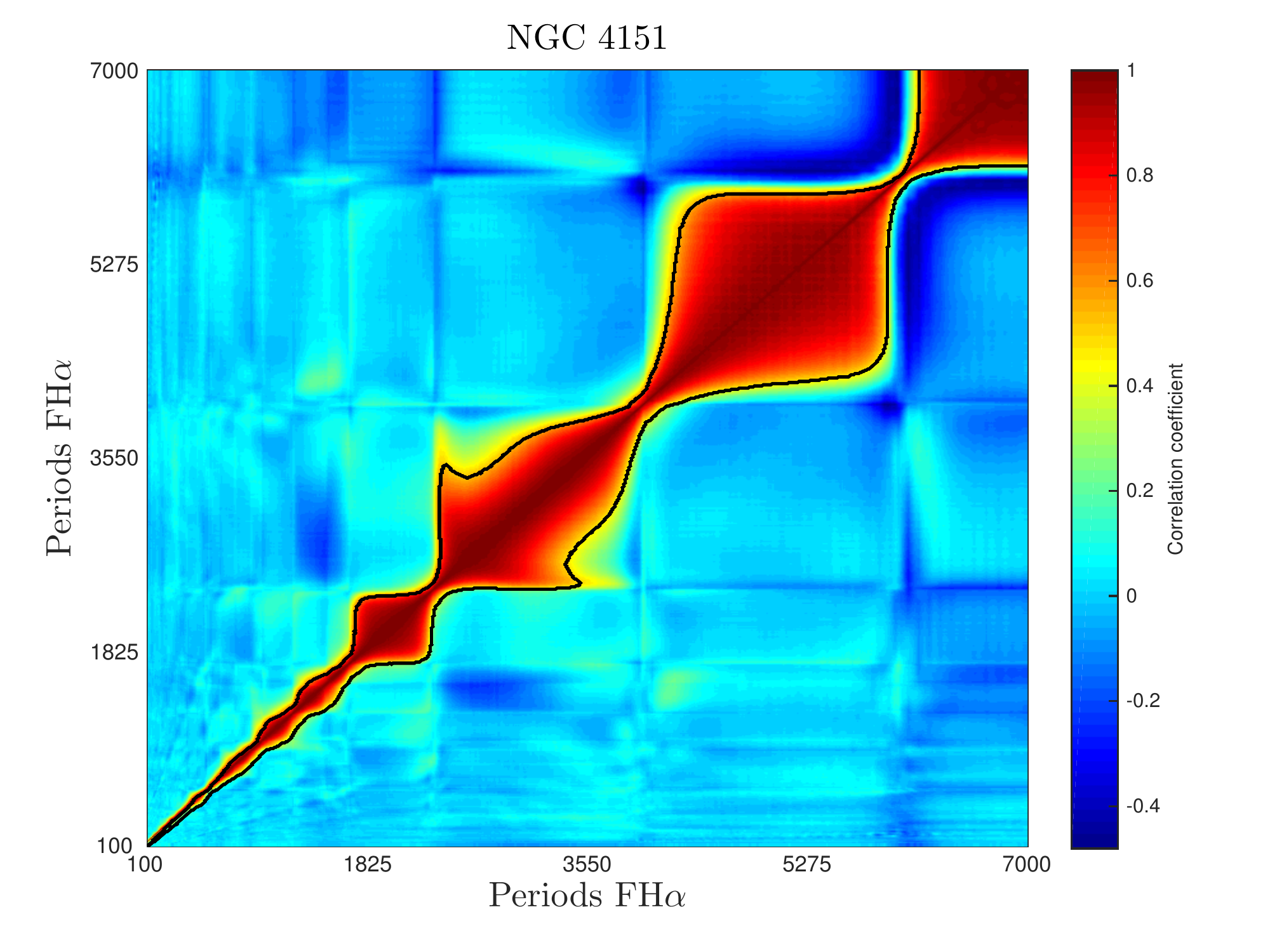}

\caption{ As in Fig. (\ref{fig:per3c}) but for  the H$\alpha$  of  NGC 4151. Note the cadence of   correlation coefficients clusters  on the diagonal.}
  \label{fig:per41}
\end{figure}

\indent \textbf{NGC 5548.} 2D correlation map of  the continuum and H$\beta$ line  shows  segregation of correlation coefficients   into single elongated  cluster  (see Fig.~\ref{fig:per55}). The cluster  is smeared between  periods of  11.75 and 14.23 yr. 
There are no off-diagonal clusters indicating synchronous coupling of period of about 13 yr (approximate center of the cluster)  in those light curves.
 \begin{figure}
    \includegraphics[width=\columnwidth]{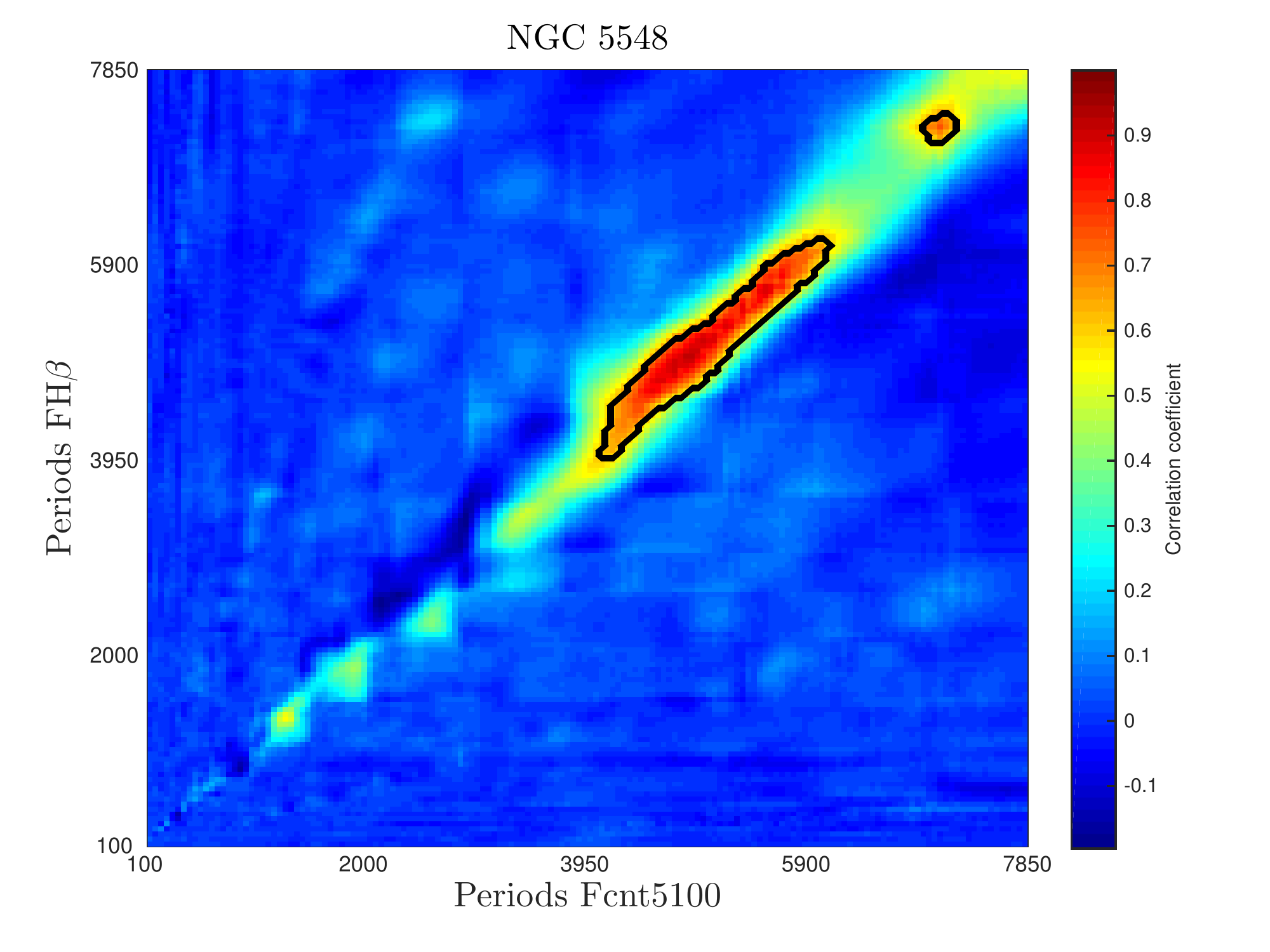}

\caption{As in Fig. (\ref{fig:per3c}) but for the  H$\beta$ and  the continuum 5100 \AA\,  of  NGC 5548.  Note 
a single elongated cluster of correlation coefficients wide about 950 days.}
  \label{fig:per55}
\end{figure}
\citet[][see their Table 8]{Pet02} reported significant changes in the H$\beta$ lag  during their 8 yr monitoring program.
We tested variability of   the time lags  between the continuum and H$\beta$ emission line  by means of windowed  Z-transformed discrete CCF  \citep[][ZDCF]{Al13}. 
Both time series were  splitted into 20 non-overlapping time segments (of different lengths ($T_{i}$))   centered on $t_{i}, i=1,...20$. Each segment encompass either convex or concave 
parts of the light curve. The  results  of the ZDCF  correlation analysis  are  summarized in  Table~\ref{tab:zdcf}.   In  Table~\ref{tab:zdcf}, the time window used for ZDCF is given in column  $win_{MJD}$. Mean,  median sampling and total points in  the  continuum and H$\beta$ line segments are given in $ \bar{P}_{cont}$, $ \widetilde{P}_{cont}$, N1 and $ \bar{P}_{line}$,  $ \widetilde{P}_{line}$, N2, respectively.  $\tau_{ML}$  is basically the time lag corresponding to  the peak of  ZDCF  $r_{\rmn max}$  nearest to zero lag, and it also has the largest maximum likelihood parameter ML. 
 
  \begin{table*}
\centering

\caption{Windowed ZDCF   parameters for  the continuum and  H$\beta$ emission line segments  of NGC 5548. The columns  are: $win_{MJD}$ is  the time window  of the curve  in Modified Julian Dates,
 $ \bar{P}_{cont}$, $ \widetilde{P}_{cont}$ and N1  stand for the mean, median  sampling periods  and the total number of  points in the continuum segments, respectively.
$ \bar{P}_{line}$,  $ \widetilde{P}_{line}$ and N2  are the mean, median 
  sampling periods and  total number of points in  H$\beta$ segments.
 $\tau_{ML}$  is  time lag corresponding to the peak of ZDCF, possessing the largest maximum likelihood parameter.
 $r_{\rmn max}$ is the peak of ZDCF,  and ML is  the value of maximum likelihood parameter.}
\label{tab:zdcf}

\begin{tabular}{cccccccccc}
\hline
 $win_{MJD}$ & & Continum  5100 \AA & & &H$\beta$    & &  & &   \\
    & $\bar{P}_{conti}$ & $\widetilde{P}_{conti}$   & N1 &    $\bar{P}_{line}$ &   $\widetilde{P}_{line}$ & N2 & $\tau_{ML}$ (days)        & r$_{max}$    &     ML  \\
\hline
 41000-42230& 26.46 &  15.80 &   32.00 & 17.84 &  11.89 &   46.00 & 25.22$^{10.66}_{-10.40}$ & 0.96$^{0.02}_{-0.03}$ & 0.53 \\

 41760-42400&  22.63 &   16.10 &   31.00 & 16.16 &11.89 &   40.00 & 25.90$^{6.58}_{-6.88}$ & 0.98$^{0.01}_{-0.02}$ & 0.66 \\
 42200-42950&24.26 & 17.40 &   32.00 & 22.88 & 15.97 &   34.00 & 21.39$^{10.54}_{-6.36}$ & 0.83$^{0.09}_{-0.12}$ & 0.50 \\
42500-43250& 22.59 &  15.70 &   32.00 &   23.00 &18.95 &   33.00 & 7.21$^{23.39}_{-6.41}$ & 0.53$^{0.18}_{-0.21}$ & 0.35 \\
43250-44000&10.16 & 3.60 &   74.00 &  15.93 &  2.06 &   47.00 & 7.76$^{28.01}_{-1.38}$ & 0.68$^{0.15}_{-0.18}$ & 0.50 \\
44100-44900&11.83 &  3.90 &   64.00 &  29.72 &21.90 &   26.00 & 22.33$^{23.26}_{-10.58}$ & 0.69$^{0.13}_{-0.16}$ & 0.27 \\
44850-45590& 10.07 &    5.00 &   74.00 & 30.62 & 24.93 &   25.00 & 20.98$^{12.61}_{-7.32}$ & 0.7$^{0.12}_{-0.14}$ & 0.45 \\
45560-47100&9.84 & 3.90 &  152.00 &  36.97 &27.08 &   41.00 & 23.09$^{0.02}_{-0.02}$ & 0.29$^{0.17}_{-0.18}$ & 0.69  \\
 47240-48000 & 5.49 &   3.00 &  138.00 &  4.99 & 2.00 &  148.00 & 12.03$^{4.36}_{-5.03}$ & 0.84$^{0.03}_{-0.04}$ & 0.88 \\
 47980-48350&8.09 &  6.00 &   57.00 &  4.85 &3.00 &   76.00 & 8.00$^{11.78}_{-1.38}$ & 0.94$^{0.03}_{-0.04}$ & 0.046  \\
48100-48440& 10.62 &  9.00 &   30.00 & 5.48 & 4.00 &   61.00 & 8.33$^{8.88}_{-1.08}$ & 0.91$^{0.04}_{-0.06}$ & 0.10 \\
 48350-48540&8.67 & 5.00 &   22.00 & 4.69 &  2.00 &   40.00 & 9.54$^{8.58}_{-3.43}$ & 0.64$^{0.18}_{-0.22}$ & 0.27 \\
48400-48900& 3.64 &   1.00 &  138.00 & 5.53 & 3.00 &   91.00 & 16.00$^{0.49}_{-7.85}$ & 0.94$^{0.02}_{-0.03}$ & 0.03 \\
48800-49260& 1.36 &  0.39 &  336.00 & 4.02 &  2.00 &  114.00 & 10.5$^{4.79}_{-0.01}$ & 0.79$^{0.04}_{-0.04}$ & 0.57 \\
49260-49700& 3.14 &  1.00 &  138.00 & 3.99 & 2.00 &  109.00 & 16.66$^{4.59}_{-0.01}$ & 0.57$^{0.09}_{-0.1}$ & 0.73 \\
49700-50100& 4.03 &  2.00 &   99.00 &4.70 & 3.00 &   83.00 & 16.73$^{5.07}_{-4.48}$ & 0.93$^{0.02}_{-0.02}$ & 0.44 \\
50000-50700&  1.69 & 0.57 &  411.00 & 3.35 & 1.00 & 207.00   & 22.02$^{11.99}_{-4.02}$ & 0.78$^{0.03}_{-0.03}$ & 0.99 \\
50700-51450& 2.72 &  0.90 &  100.00 & 4.27 & 1.00 &  169.00 & 22.77$^{5.50}_{-7.02}$ & 0.87$^{0.05}_{-0.06}$ & 0.34 \\
51450-52207& 3.29 &  0.99 &  219.00 & 8.25 &   3.00 &   88.00 & 17.04$^{12.95}_{-5.39}$ & 0.55$^{0.09}_{-0.09}$ & 0.57 \\
52207-56660& 31.26 & 2.00 &  121.00 & 56.08 &  22.00 &   78.00 & 28.47$^{0.01}_{-0.01}$ & 0.69$^{0.08}_{-0.09}$ & 0.54  \\
\hline

\end{tabular}
\end{table*}

  In  Fig. ~\ref{fig:blr}. we show $\tau_{ML}$  as a function of  the centers of non-overlapping windows presented in Table \ref{tab:zdcf}.
Even the size of set of  time lags  may be small, we fit a sinusoid   (see  Fig. ~\ref{fig:blr}), which   assumes the  periodicity of  $\sim$ 4600 days. 
This function  is intended to guide the eye along the data points, not to state that time lags behave
 necessarily  periodic.
The fitted curve does not  pass through all the points, but we note that about the half  of the points are   on each side of the curve. 
It may be worthwhile to mention that   there are reasons to expect
periodicity in the time lag evolution curve, since BLR size varies with the mean optical luminosity \citep{Lu16}.
 By contrast, the estimated sinusoidal fit to the observed light curve (Table~\ref{tab:resultfit})  is of poor quality, because data are highly volatile, with frequent peaks and troughs, which are often quite sharp (see  Fig. ~\ref{fig:sinusfit}).

\begin{figure}
    \includegraphics[width=\columnwidth]{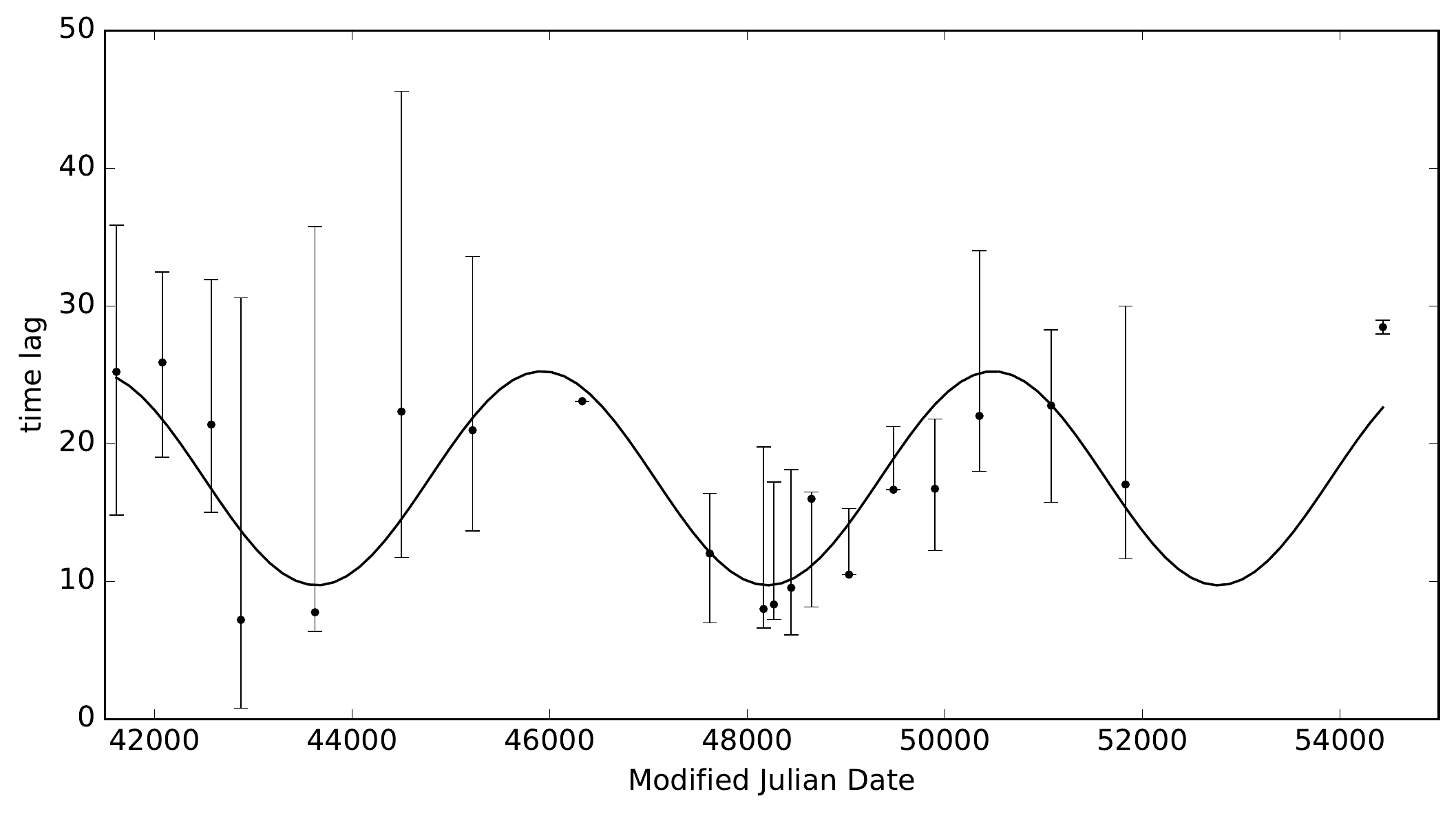}
\caption{ Time evolution of the size of BLR of NGC 5548 compiled from the data given in Table~\ref{tab:zdcf}. 
It shows  the estimated  time lags between the continuum and H$\beta$ emission line from windowed ZDCF against  the center of time windows. Superimposed solid line is the best-fitting of a sinusoid of
period $\sim$ 4600 days.}
 \label{fig:blr}
\end{figure} 
 \indent \textbf{E1821+643.}
 Using the data for  all light curves, our method yielded periods  at  $\sim$ 12 yr, $\sim$ 6 yr and $\sim$ 5 yr (see Fig. ~\ref{fig:pere}). It is interesting that periodicities have been revealed in the case of H$\beta$ emission line. One can note that the period of  $\sim$12 yr is about  twice  of  $\sim$ 6 yr period. The correlation maps of E1821+643  show a tail of smaller periodicities disconnected from the prominent  cadence of clusters of  longer periods, but it is  globally similar to the  topology of periods  found  in NGC 4151.
 { Three sinusoids fit is of comparable quality to the case of NGC 4151 (see Table~\ref{tab:resultfit}).
 
 \begin{figure*}
\centering
\begin{tabular}{cc}
    \includegraphics[width=0.47\linewidth]{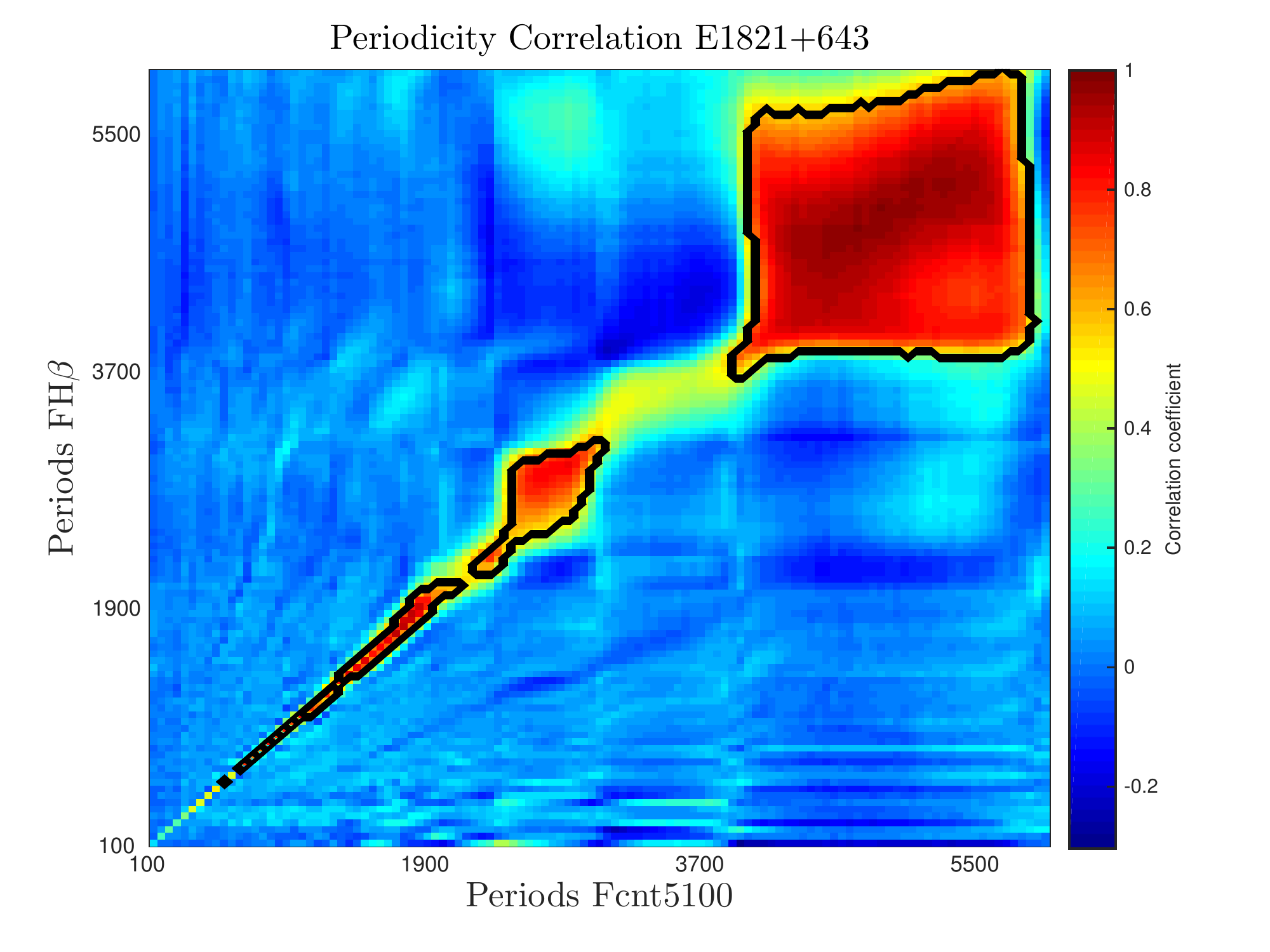}&
    \includegraphics[width=0.47\linewidth]{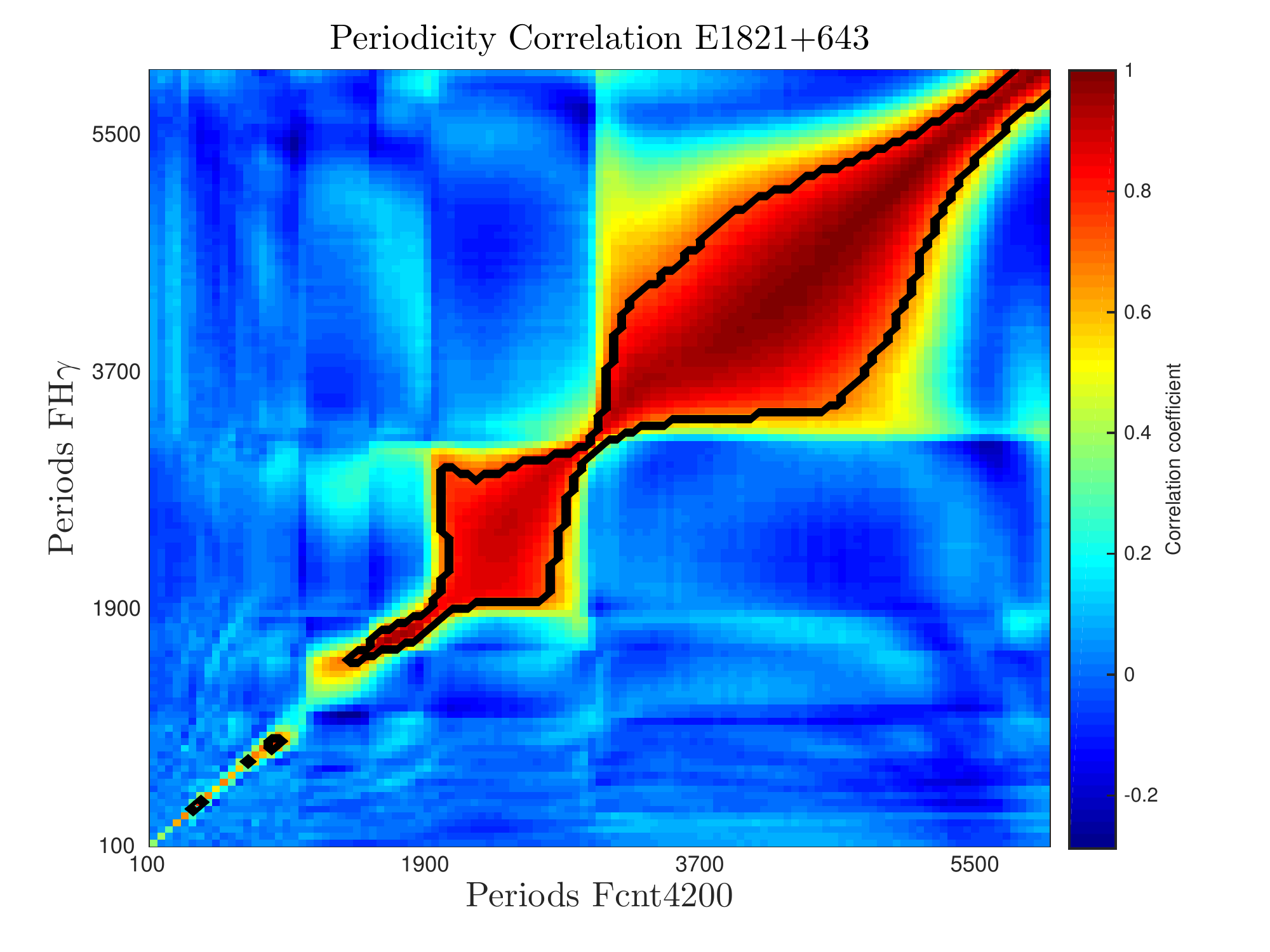}\\ [2\tabcolsep]
 \end{tabular}
\caption{As in Fig. (\ref{fig:per3c}) but for all  light curves of E1821+643. Left panel: continuum  5100 \AA \, vs.  H$\beta$ emission line.  Right panel: continuum 4200 \AA\,  and H$\gamma$  emission line. Note that on both panels the tail of smaller periodicities is disconnected from the prominent  correlation clusters. }
  \label{fig:pere}
\end{figure*}

\begin{figure}
\includegraphics[width=0.5\textwidth]{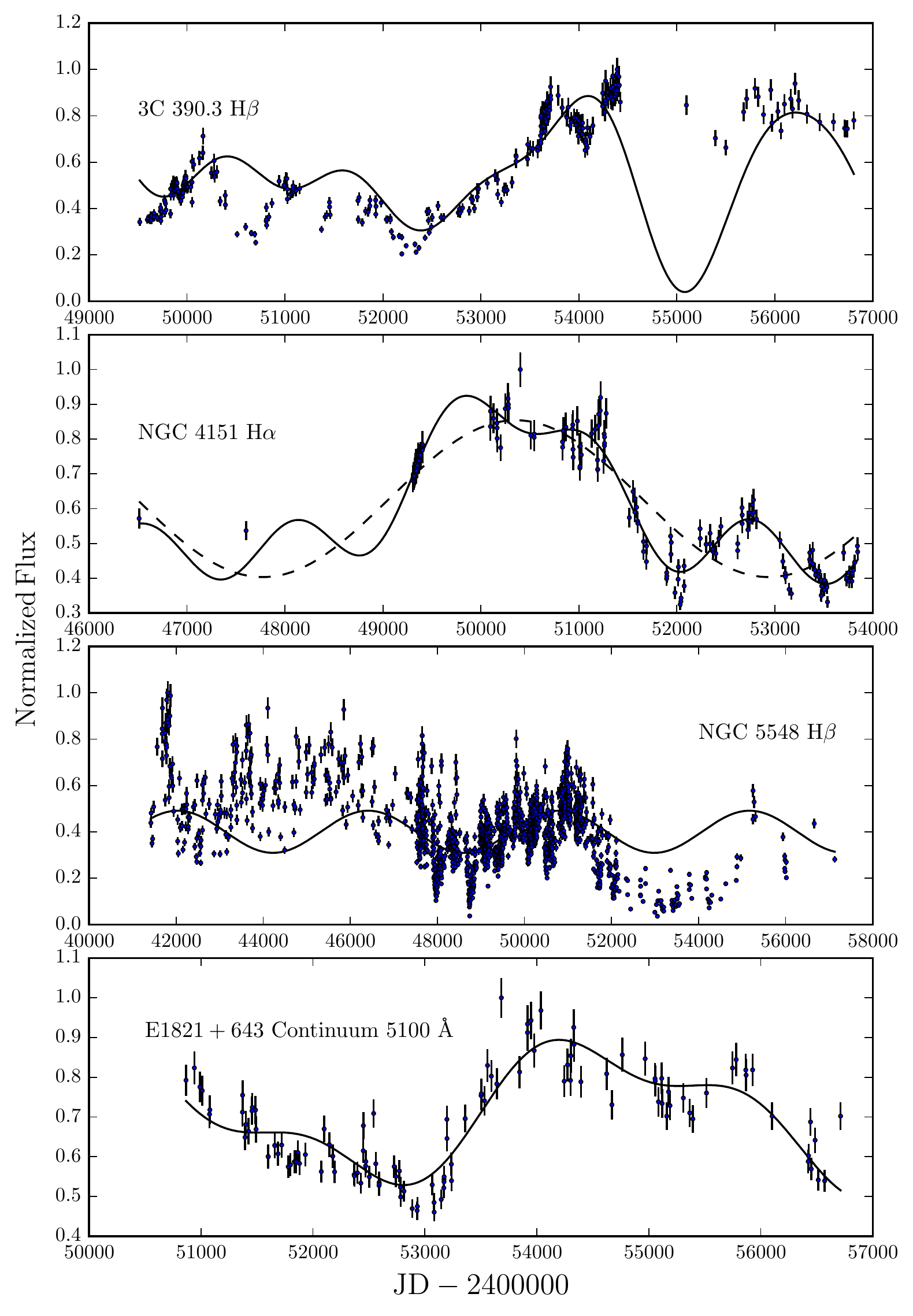}
\caption{  Best-fitting of  multisinusoidal models to the observed light curves. Normalized fluxes  are represented by dots with error bars whereas models with solid lines. From top to bottom: sum of 5 sinusoids  for  H$\beta$ line of 3C 390.3;  sum of 3 sinusoids (solid line) and one sinusoid corresponding to the largest period (dashed line) for   H$\alpha$ line of  NGC4151; one sinusoid for  H$\beta$ line of NGC 5548 and sum of  3 sinusoids for  the  continuum 5100\, \AA\,  of E1821+643. Note that sinusoid model for Arp 102B is missing, since we could not determine any periodicity in its light curves.}
   \label{fig:sinusfit}
\end{figure}

\section{Discussion}
In general, the derived optical periodic variability of our  AGN sample  is  in agreement with  previous investigations.
However,  our hybrid method revealed  specific  topology of  correlation  between  oscillatory patterns in the light curves that were missed in some other
works. Particularly, we can more clearly separate objects according their oscillatory patterns. 
In the following, we discuss these oscillatory patterns from the point of view of 
abstract models of coupled oscillators.  For each source we  constructed such model which resembles detected oscillatory patterns.

\noindent  \textbf{Double-peaked emitters 3C 390.3 and Arp 102B.} 
  Our   hybrid method  made distinction  between these  two objects based on  the presence (absence) of  underlying periodic variability, despite that the differences in their  shape of  spectral lines   are subtle.  
 2D (auto)correlation maps of Arp 102B  have simple linear shape, but for  3C 390.3  a manifold of substructures is superimposed.
Namely, our analysis has  shown the  fluctuations in the interactions  between the oscillations in the  continuum and the  emission lines of 3C 390.3 due to substantially different levels of correlation polarity.
The clusters of correlations  are divided by gaps of very low (or no) correlations, and the islands of correlations switch their polarity (between positive and negative values).  
This  can be seen  as an evidence that  there  exists an activation process which  reduces correlation coefficients (from positive values up to zero).  After,   different  relationship is  evoked  between the  periodicities in light curves (e.g. negative correlations). 
 The break off of the correlation band (on diagonal)  may be due to the  change of variability in the light curves  or due to the modulatory effects of some unknown background factors, which may have certain  cycles. 
We applied the hybrid  method to a series of  artificially non-linear and non-stationary oscillatory signals simulated by   Eq. ~(\ref{eq:eq3}) and ~(\ref{eq:eq4}) 
with unidirectional and bidirectional coupling. The degree of non-linearity/asymmetry is controlled by the ratio (differing from 1) of the periods (frequencies) of the sinusoids.
 To  introduce non-stationary mechanism,  we assign a random values between 0 and 10 to the amplitudes of  signals, while they were fixed over trial realization, adding the red noise.
2D correlation map of   bidirectional coupling of three oscillatory processes
(see Fig.~ \ref{fig:sim3c}) is almost matching the  map of the real light curves.

\begin{figure*}
\centering
\begin{tabular}{cc}
    \includegraphics[width=0.47\linewidth]{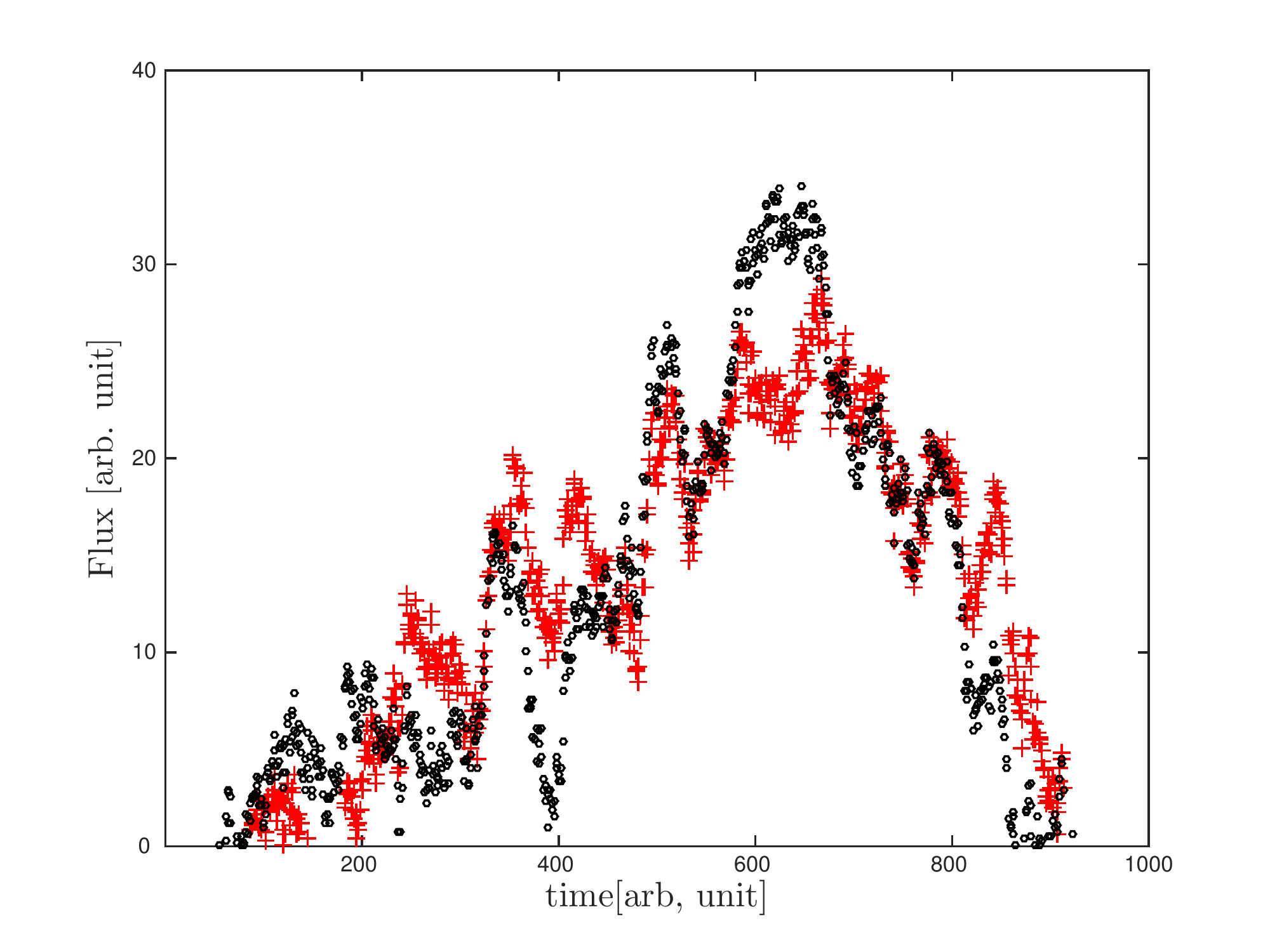}&
    \includegraphics[width=0.47\linewidth]{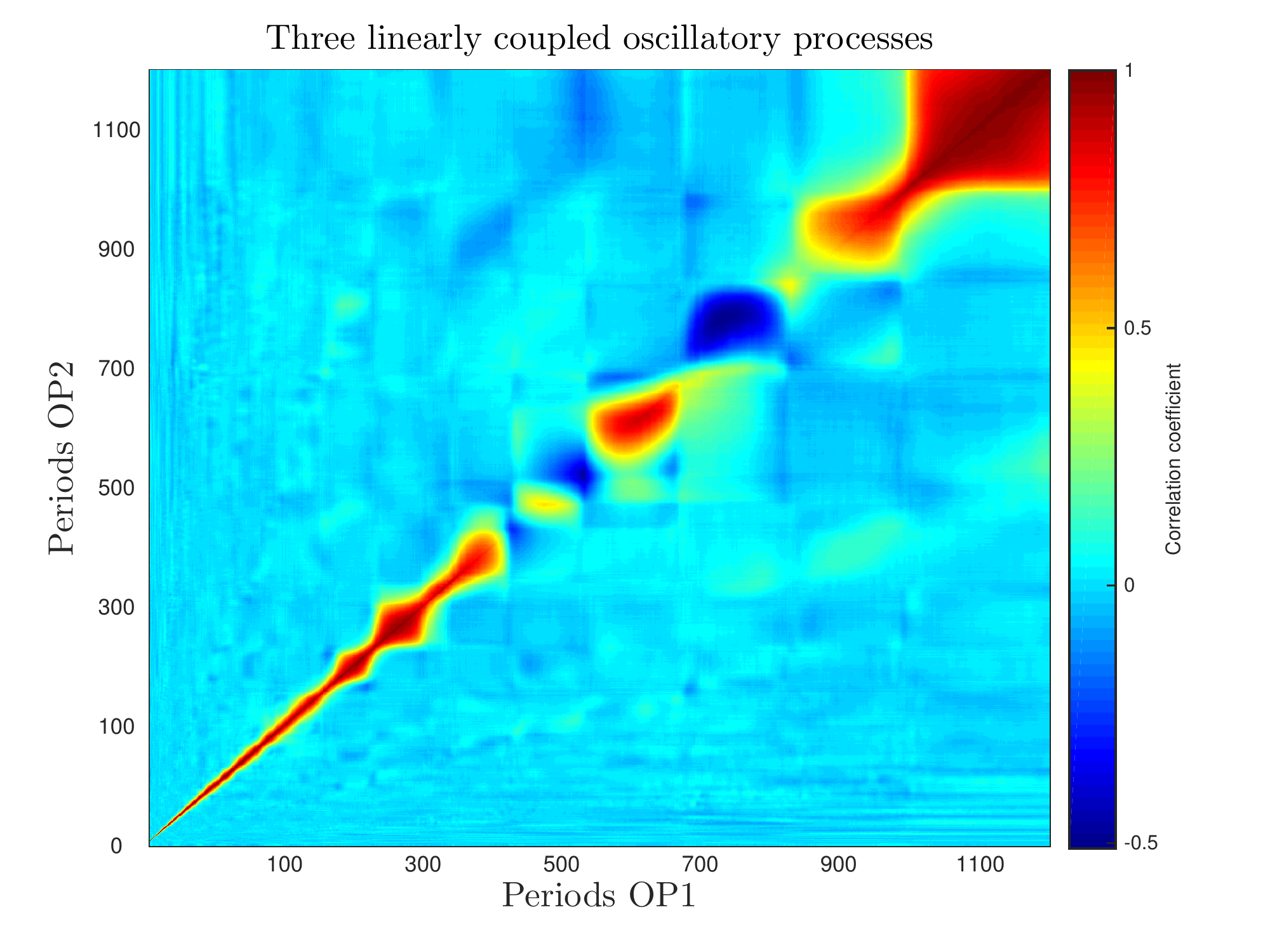}\\ [2\tabcolsep]
 \end{tabular}
\caption{Simulation of   bidirectional coupled three oscillators network for the case of 3C 390.3. Left: Random realization  of Eq.~(\ref{eq:eq4}) 
  form two time series (black is $U_{a}=OP1$, and red is $U_{c}=OP2$) of amplitudes $A=1.954, B=1.729,  C=2.357$, phase $\phi=\phi_{1}=2.359\,\, \mathrm{rad}$ coupling strengths  $cp_{a\rightarrow b}=0.7,cp_{a\rightarrow c}=0.5, cp_{b\rightarrow a}=0.2, cp_{c\rightarrow a}=0.3$,  frequencies  $f_{a}=\frac{1}{1000}, f_{b}=\frac{1}{300}, f_{c}=\frac{1}{100}$,  and  time delay  of 100 arbitrary chosen time unites. Right: corresponding 2D correlation map, which  clearly shows
   three clusters related to fundamental periods as well as clusters of negative correlation.}
  \label{fig:sim3c}
\end{figure*}

\begin{figure}
    \includegraphics[width=\columnwidth]{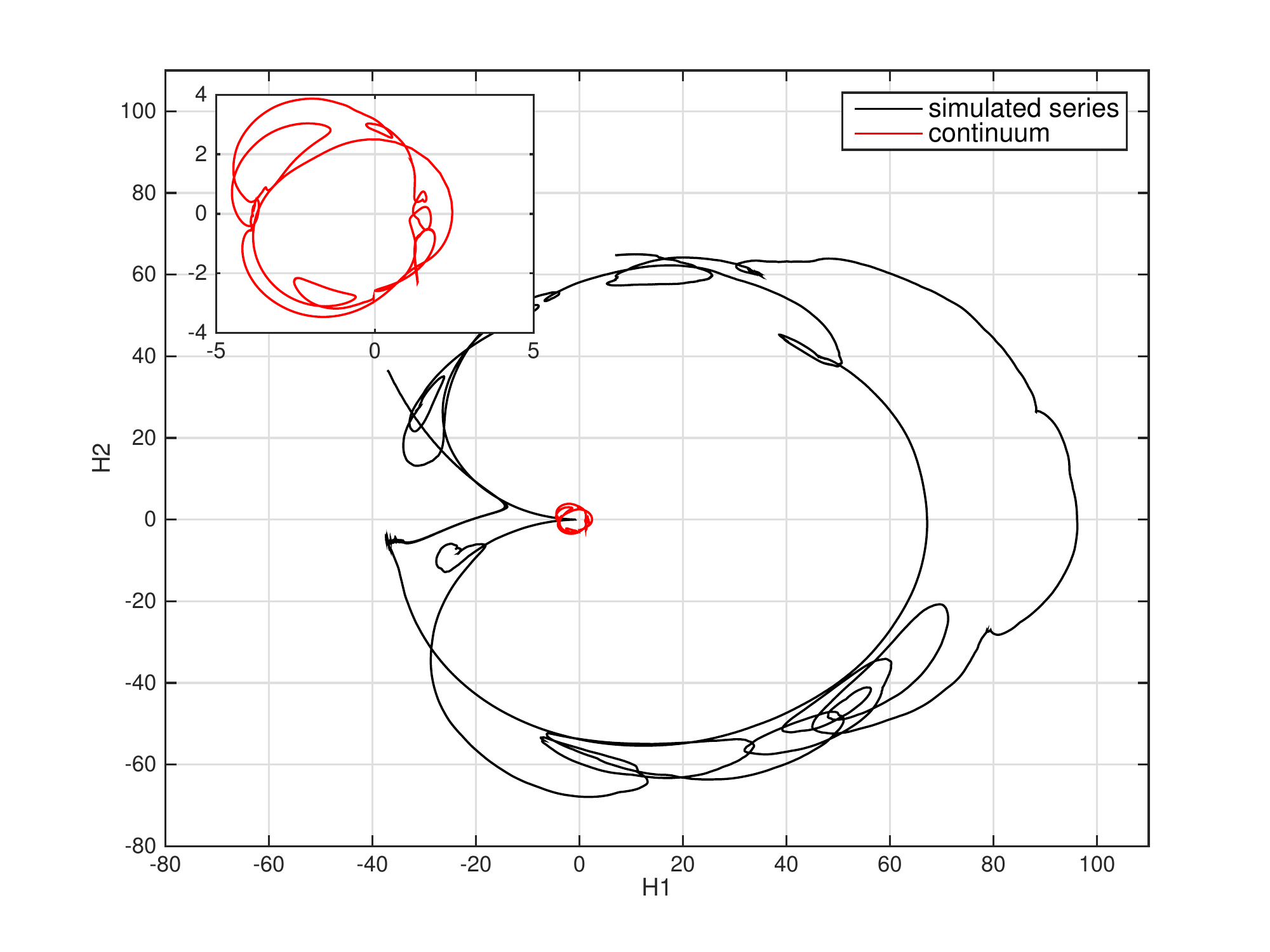}
\caption{Comparison of the phase trajectories  between the continuum of 3C 390.3 and simulated curve  OP1 from oscillatory network model in Fig. ~\ref{fig:sim3c}. H1 is  related to  the time series  itself (real part),
 while H2 is imaginary part  of analytical signal  (see Eq.~(\ref{eq:eq6})). The inset and main plot are very topologically similar (beside a phase shift), showing somewhat distorted ovoid curve.}
 \label{fig:ph3c}
\end{figure}

 We could not produce correlation image with negative 'correlation islands'  if the time delays between processes are different. We note that time delays between the  continuum and
the  H$\alpha$ and H$\beta$ lines are $\sim$ 120 and 95 days, respectively,  as reported  by \citet{Sh10a}. 
Based on this, one can see the  importance of value of  time lag for  expression of negative link between oscillations which implies that physical places of oscillations sources  are somehow functionally related to each other.  
 For example, such relationship can be 'hot spot'. \citet{Jov10} showed that  two large amplitude outbursts of the H$\beta$ line observed between 1995 and 1999 in 3C 390.3 could be explained by successive occurrences of two bright spots in the accretion disc.
The phase plots in Fig.~ \ref{fig:ph3c} illustrate comparison between behaviors of the continuum and one of  simulated curves from oscillatory network model for 3C 390.3.
 Both curves  show a specific  mode of dynamics, generating a major loop (large amplitude oscillation) and the formation of  secondary loops
 (small amplitude oscillations). 
This confirms   that the signal of oscillations  is not a monocomponent
but a multi-periodic. Both trajectories  cover evenly  the section of phase space   shown here. Particularly, there are no single regulation points at which multiple loops of  both  trajectories 
can intersect.  The main loops of  trajectories are of different diameters due to different amplitudes of real and simulated curves. The secondary loops   can be regarded as a hidden attractor.

A specific topology of Arp 102B correlation maps indicates  either that periodicities are absent from light curves or that coupling between oscillatory processes  is weak.
We have created about 100 pairs of artificial curves consisting of 2000 points based on  two linearly and weakly coupled oscillators (see Eq. (\ref{eq:eq3})).
For reference,  the most realistic realization of simulation producing the  correlation image matching the linear oscillatory patterns of observed light curves  is presented in Fig.  \ref{fig:simarp}. 
\begin{figure*}
\centering
\begin{tabular}{cc}
    \includegraphics[width=0.47\linewidth]{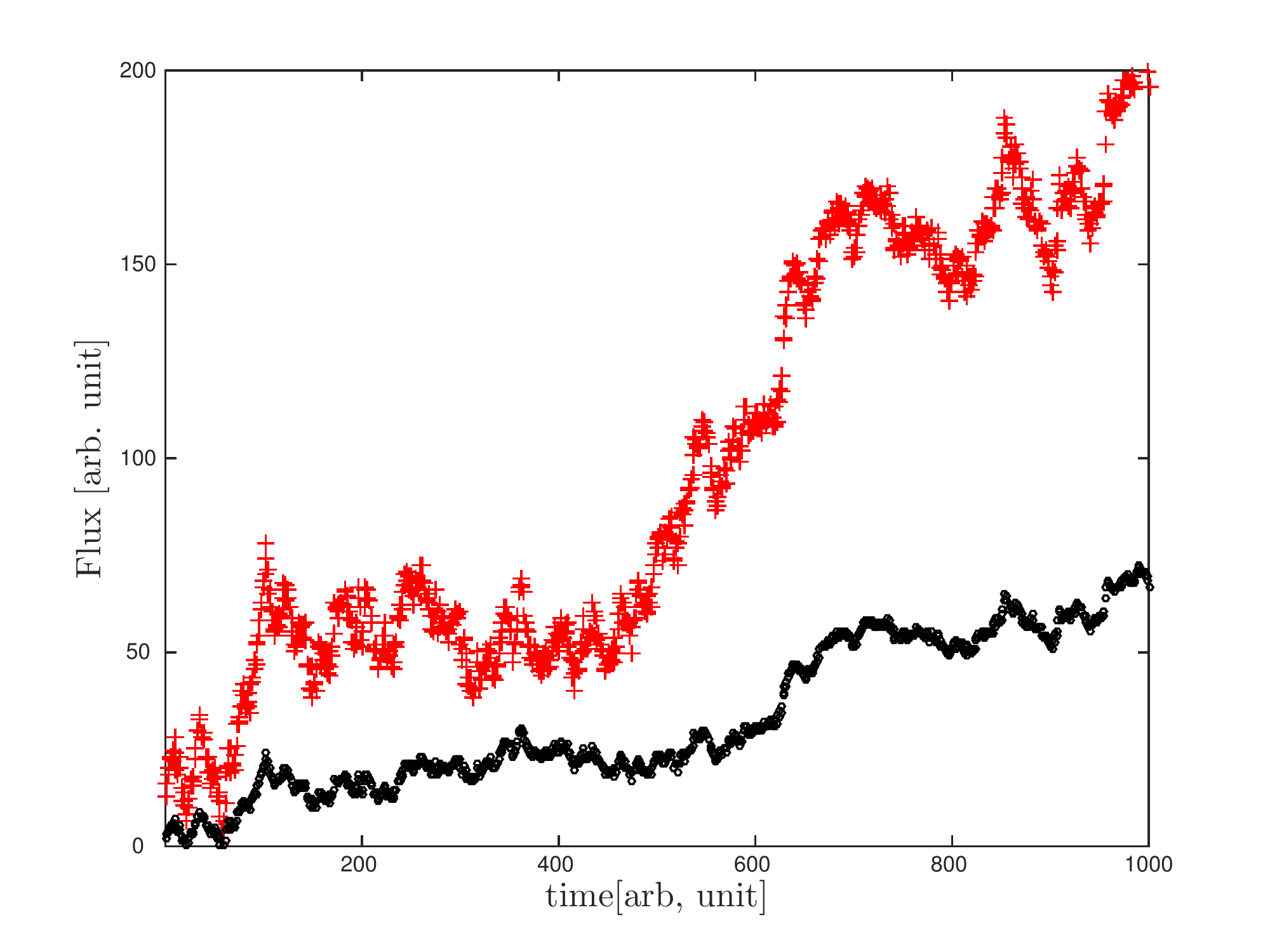}&
    \includegraphics[width=0.47\linewidth]{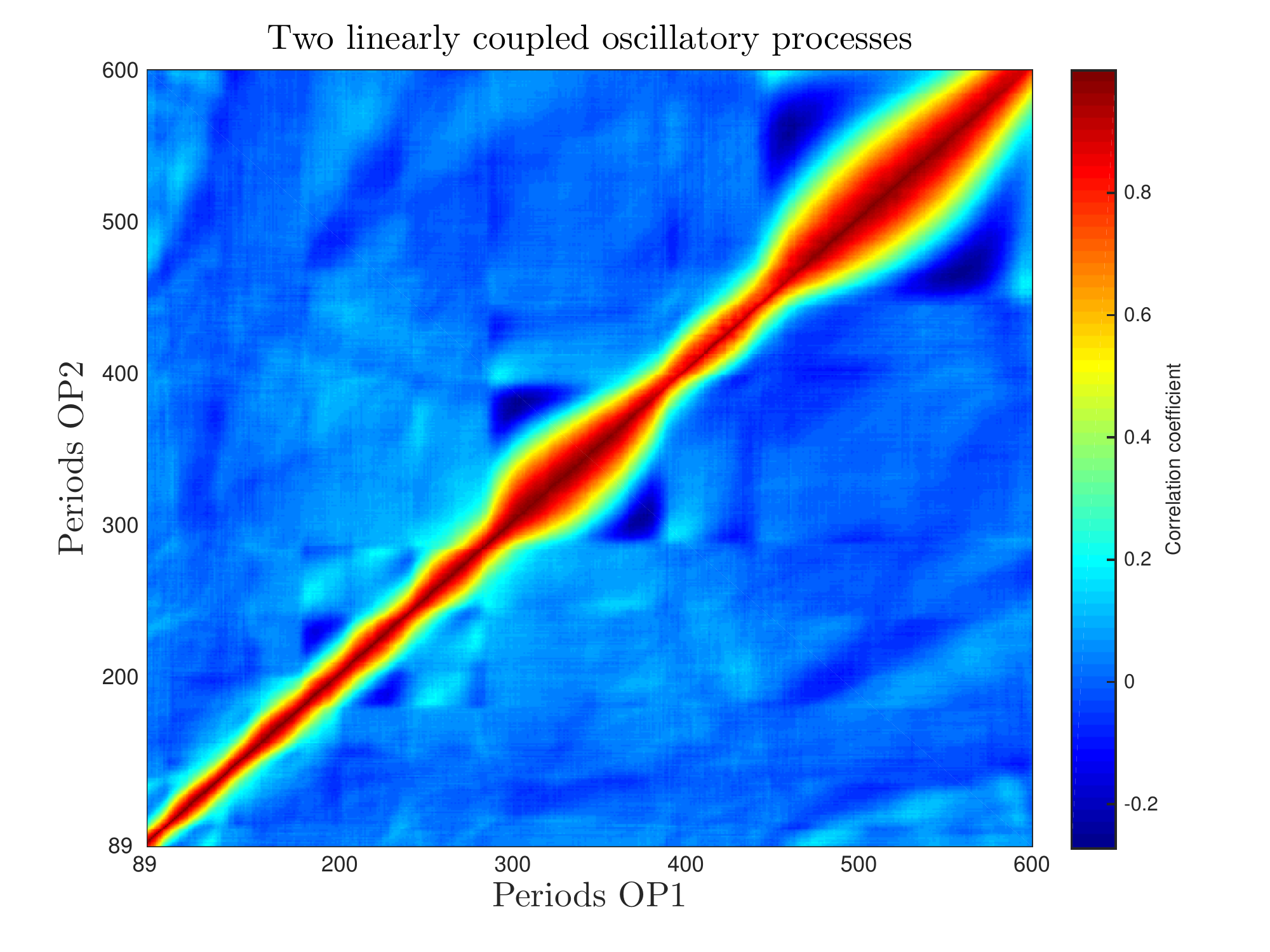}\\  [2\tabcolsep]
 \end{tabular}
\caption{Simulation of   two bidirectional coupled oscillators  for the case of  Arp 102B. Left:  
Random realization  of  Eq.~(\ref{eq:eq3}) 
  form two time series (black is $U_{a}=OP1$, and red is $U_{b}=OP2$) of amptitudes
 $A=5.29, B=1.99$, phase $\phi=0.4174~ \mathrm{rad}$, coupling strengths $cp _{a\rightarrow b}=0.4, cp _{b\rightarrow a}=0.2$, time delay is 100 and periods are 500 and 300  arbitrarily chosen time units.  Right: corresponding   2D correlation map.}
  \label{fig:simarp}
\end{figure*}

In comparison with  3C 390.3, it seems that with  decreased coupling of oscillators of Arp 102B,  the
clusters merge, decreasing   the variance of the distribution of oscillation periods, which  eventually
results  in a   system where oscillations are absent.
There is a very close match between phase portraits (see Fig. \ref{fig:pharp}) of observed H$\alpha$ and simulated curve demonstrating that the model of weakly  coupled oscillators is capable of capturing the dynamics
of this object. Note that phase trajectories  appear to cross over themselves. Actual trajectories in 3D space (adding the time as third coordinate) are spirals and do not cross. 
The apparent crossings are due to projection on 2D phase plane.
They are similar in topological sense without forming the smaller loops,  but they are  non closed,  indicating either of weak coupling between oscillators or absence of periodicity.

\begin{figure}
    \includegraphics[width=\columnwidth]{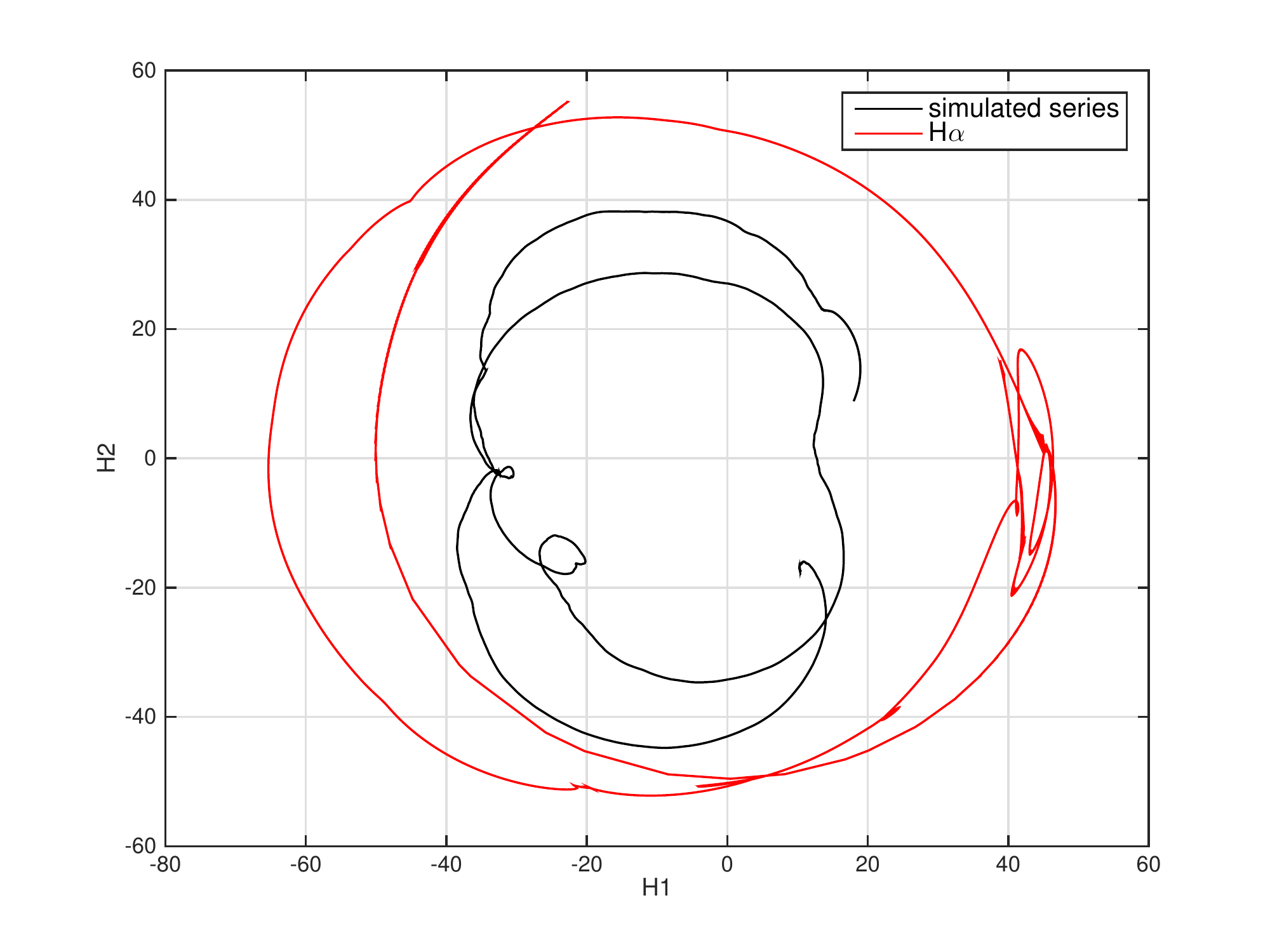}
\caption{As in Fig. ~\ref{fig:ph3c} but for  the H$\alpha$ line of Arp 102B and simulated  OP1 curve from oscillatory  model (see Fig. \ref{fig:simarp}). 
 Both curves are similar and non-closed, indicating either weak coupling or absence of periodicity. They appear to intersect themselves due to projection on 2D phase space.} 
  \label{fig:pharp}
\end{figure}

 The lack of   oscillatory patterns  favor non binary BLR hypothesis for this object.
This is in  line with recent study of  \citet{Liu16}, where is  reported  that the estimated Arp 102B mass of $1.1\times 10^{8} M{\odot}$ \citep{Sh13} is far less than $10^{12}M{\odot}$ which is  obtained under binary black hole
assumption.
The topology of  correlation map of oscillatory patterns of  Arp 102B which we found, can be best  explained  in the context of the process causing  gradual variations which are stable over long monitoring period.
In spite of  both objects 3C 390.3  and Arp 102B  being classified as double-peaked emitters, we have shown that underlying  topology of their   oscillations mechanisms    are quite different, suggesting
 different physical background \citep[see datailed discussion in][]{Pop11,Pop14}. 

\noindent  \textbf{Supermassive binary black hole candidates NGC 4151, NGC 5548 and E1821+643.} 
\noindent  Again, our hybrid method discerns the oscillatory dynamics of  studied  objects.
 Cadenced topology of three detected  periods in the H$\alpha$ line of NGC 4151 led us to suspect that  periodic signals can be non-linearly coupled.
  We simulated such coupled oscillatory system using following equation 

\begin{equation}
\begin{aligned}
U_{a}(t)={}& A(t)\cdot \sin (2\pi f_{a} t +\phi)+cp_{b\rightarrow a}\cdot\\
                    & B(t)\cdot \sin(2\pi f_{b} t+2\pi f_{b}\tau) + W(t)\\
U_{b}(t)= {}&B(t)\cdot \sin (2\pi f_{b} t)+cp_{a\rightarrow b}\cdot \\
                    &{U_{a}(t)}^{2}+W(t)
\end{aligned}
\label{eq:eq8}
\end{equation}

\noindent  where the non-linear coupling is introduced by squared term ${U_{a}(t)}^{2}$.
Simulated curves consists of sum and multiple of base sinus signals of  periods of 500 and 300 arbitrary chosen time units. 
As a consequence,  periods of  $2*500,2*300, 500, 300$ are accompanied with an interference patterns $500+300,500-300$ (right plot in Fig. \ref{fig:sim41}). Comparing this scenario with autocorrelation of periods in H$\alpha$ (see Fig. \ref{fig:per41}), the largest period of 13.76 yr can be interpreted as interference pattern (i.e. sum) of two smaller periods of 5.44 and 8.33 yr.

\begin{figure*}
\begin{tabular}{cc}
    \includegraphics[width=0.47\linewidth]{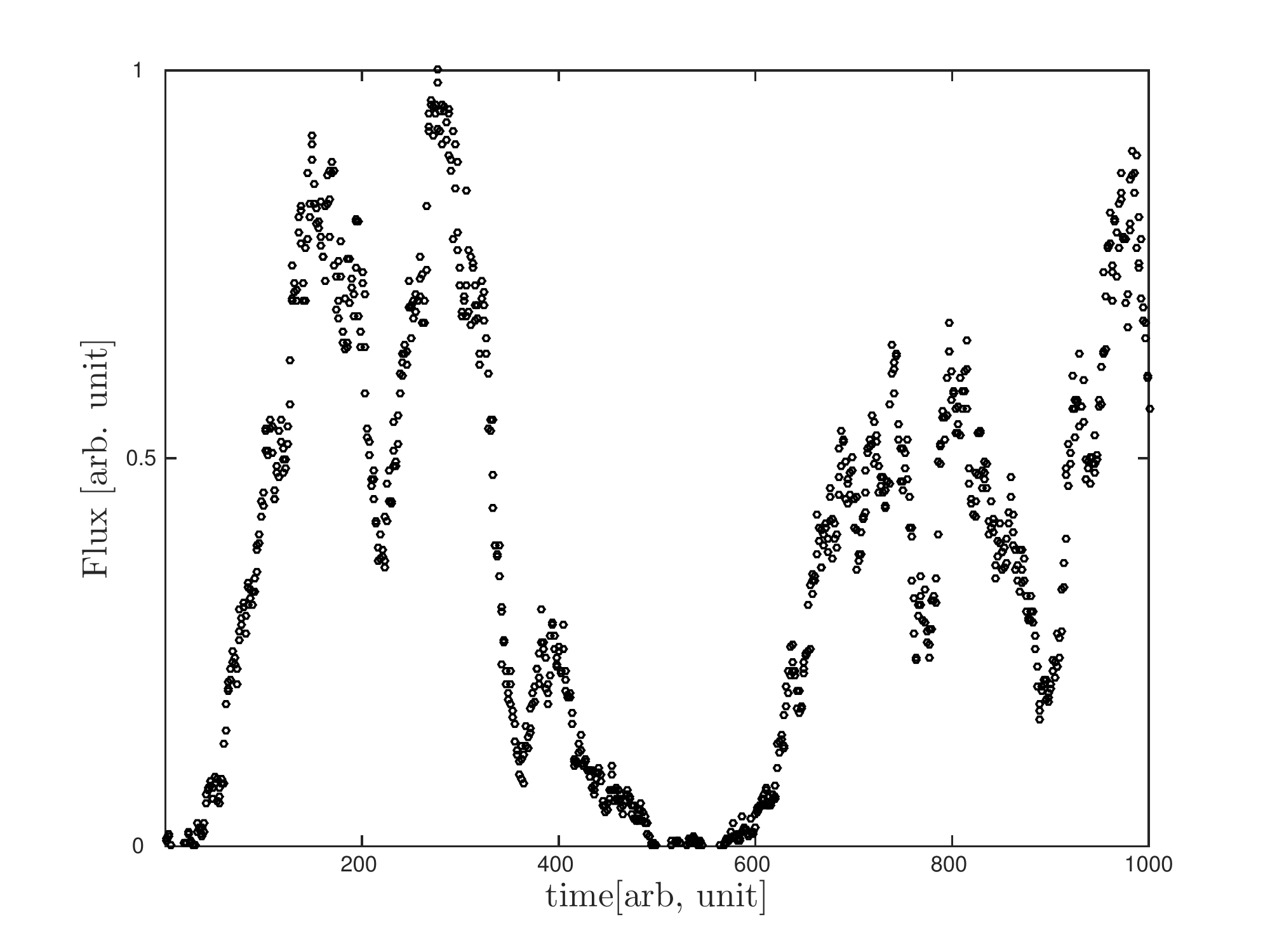}&
    \includegraphics[width=0.47\linewidth]{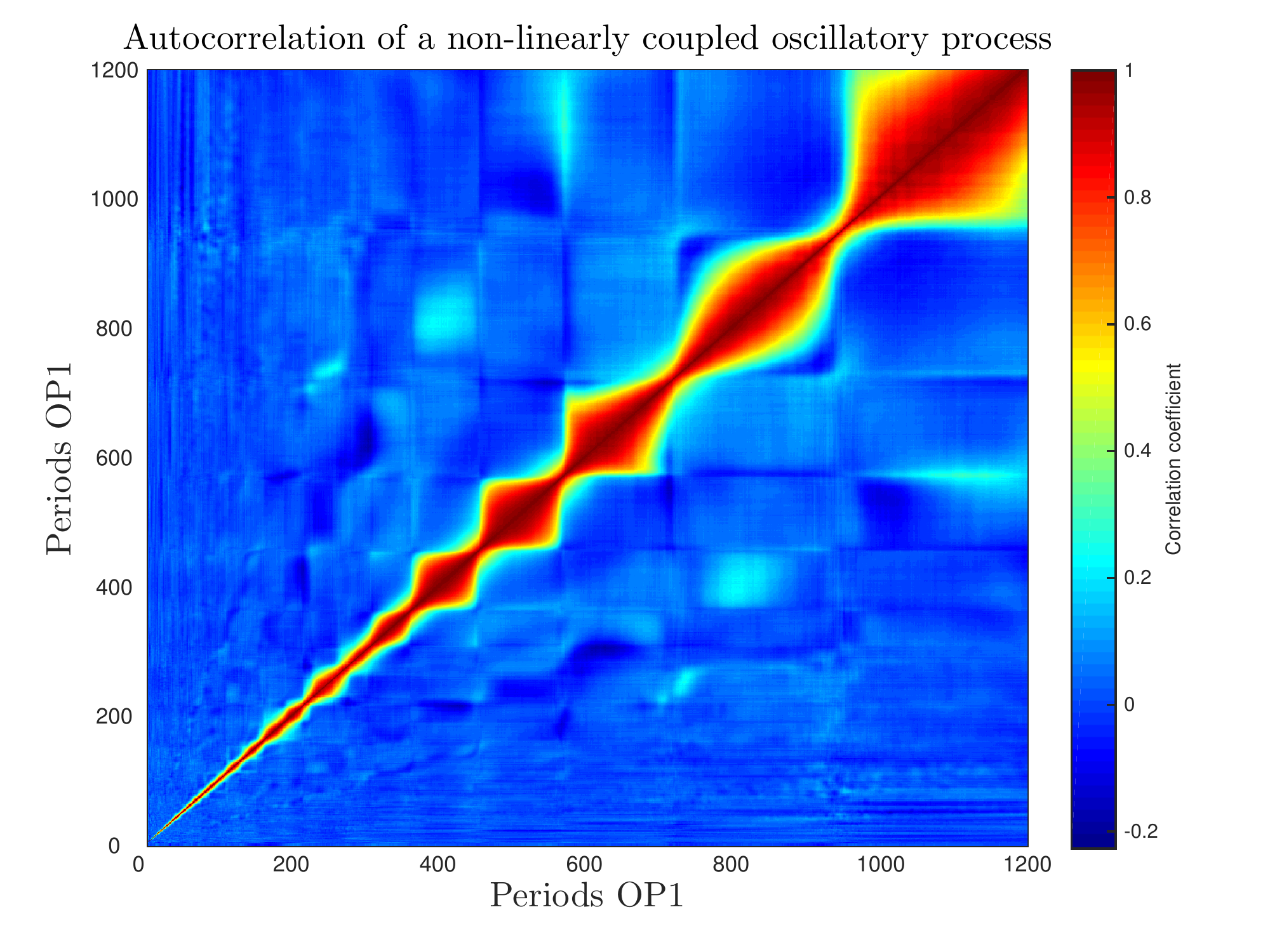}\\  [2\tabcolsep]
 \end{tabular}
\caption{Simulation of   two bidirectional coupled oscillators  for the case of  NGC 4151. Left:  
Random realization  of  Eq.~(\ref{eq:eq8}) 
  form two time series (black is $U_{a}=OP1$, and red is $U_{b}=OP2$) of amplitudes   $A=6.09, B=1.04$, phase $\phi=2.2\, \,  rad$, coupling strengths $ cp _{a\rightarrow b}=0.7, cp _{b\rightarrow a}=0.6$, periods are 500, 300 and   time delay is 100 arbitrarily chosen time units. Note the similarity of  sharpness of this signal 'bursts' with features in the observed light curves. Right:  corresponding  2D correlation map .}
  \label{fig:sim41}
\end{figure*}

\begin{figure}
\includegraphics[width=\columnwidth]{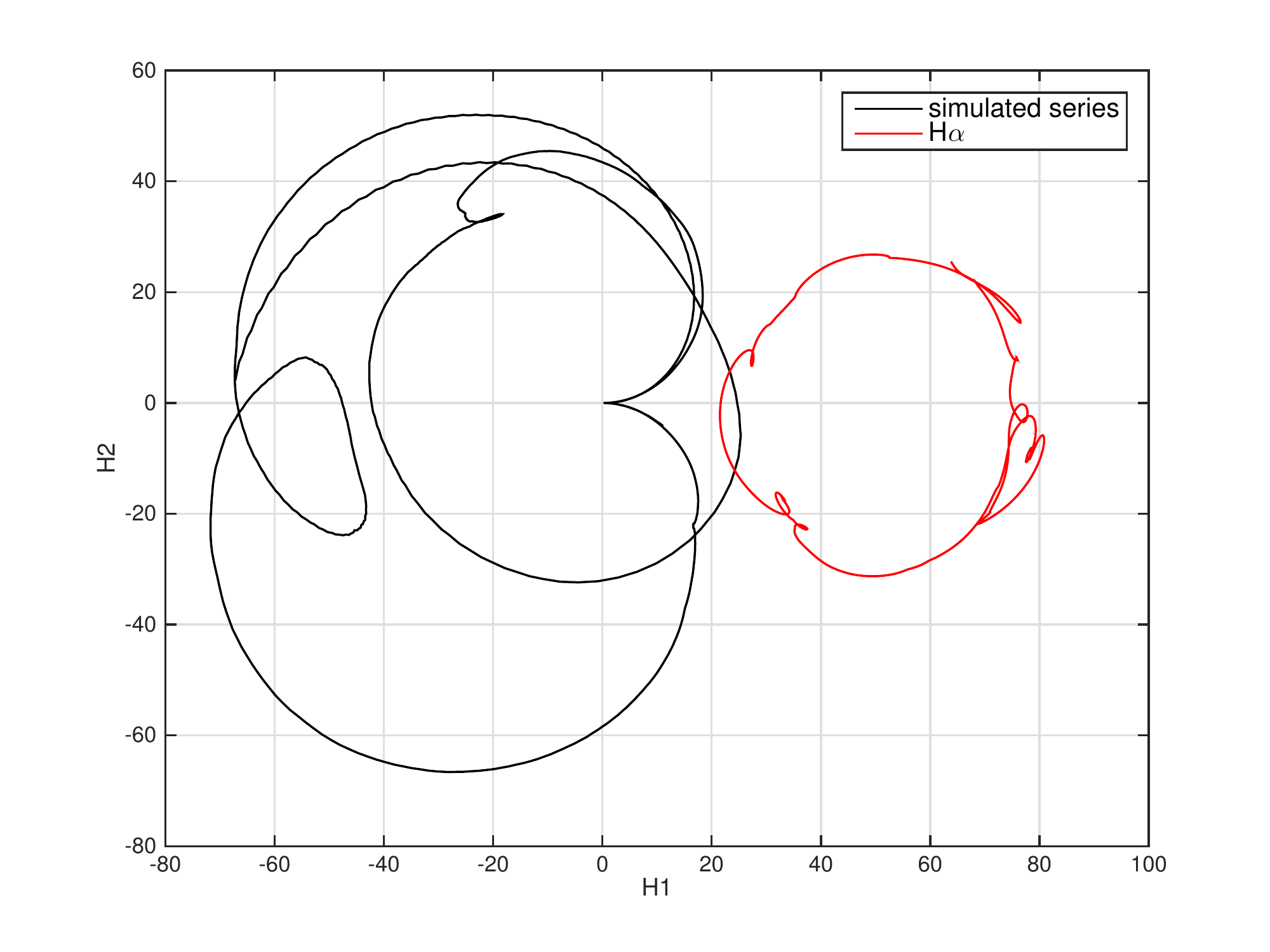}
\caption{As in Fig. ~\ref{fig:ph3c} but for  the H$\alpha$ line of NGC 4151 and  simulated OP1 curve described by Eq. (\ref{eq:eq8}) with parameter values as in  (see Fig. \ref{fig:sim41}).
Note that phase curve of H$\alpha$ line is shifted by + 50 units on x axis  for a better view.} 
   \label{fig:ph41}
\end{figure}

If presented, off diagonal correlations clusters among different periodicities  are  an indicator of  an asynchronous coupling.
Note  that the correlation map of NGC 4151 does not have such topology, which implies that the  coupling between periodicities in this object  is synchronous. One way to enhance synchronization among non-linearly coupled  oscillators is by  increasing the coupling strength between them.  Due to this, in the model
both  coupling strength  were set  high (see  Fig. \ref{fig:sim41}).

The phase portrait  of the  H$\alpha$ line  (see Fig. \ref{fig:ph41}) shows formation of  three  smaller  loops outside of the main loop, confirming  mult-periodic oscillations found in  2D correlation map. 
Moreover, from the same figure it  can be seen   that the numerical simulation shows  similar  phase trajectory, except the size and the  position of a heart-like smaller loop.
 This is a consequence of  higher  amplitude of  associated component of oscillations in the model.
On both curves the major loop corresponds to the relatively stable periodic motion of larger amplitude, while smaller loops correspond to the oscillations of smaller amplitudes.

The  variability of  the continuum and  H$\beta$ line of NGC 5548 over 30 year \citep{Ser07} and over 40 years \citep{Bon16}, has revealed sharp peaks which are similar to the case of NGC 4151, suggesting presence of non-linearly coupled oscillations. Moreover, cross correlation functions  (CCF) between the continuum and H$\beta$ line over long time span estimated by  \cite[][see their Fig. 4]{Ser07}  are centered and non deformed, suggesting synchronization between  those two light curves.\footnote{\label{foot:ccf}That is a consequence of analogy between the CCF and inner vector product. The  the inner product measures the angle between two vectors, so CCF can give a sense of the level of the synchronization of two signals.}
  Namely, when the system is asynchronous, the magnitude of the typical CCF values  are  smaller and   asymmetric (i.e. shifted), while  synchronous system exhibits larger and centered  CCF values.   Fig. (\ref{fig:simcross})  illustrates  the effects of asynchronization between  two sine waves with frequencies of  $\frac{2\pi}{10}$ on their cross correlation function. The starting phase of one sine wave is 0, while the starting phase of the other sine wave is  $-\pi$ and  both signals  are contaminated with  red  noise. The calculated CCF is clearly shifted and its values are reduced.  
   \begin{figure}
    \includegraphics[width=\columnwidth]{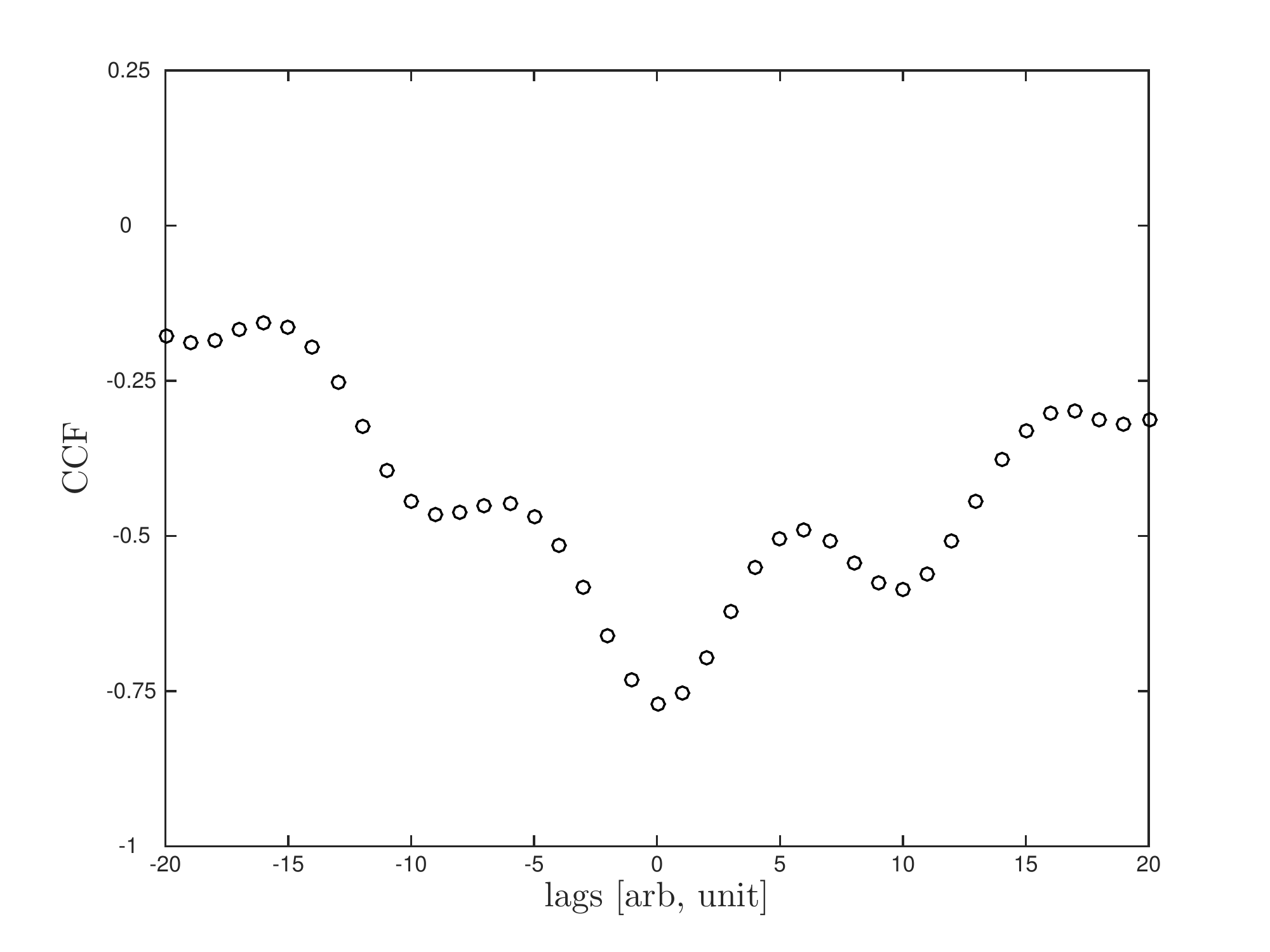}

\caption{Simulation showing how the cross correlation function is affected by asynchronyzation between two sine waves.  Both signals oscillates with  the same frequency of  $\frac{2\pi}{10}$ but   with different starting phases. Both signals are contaminated with red noise.}
  \label{fig:simcross}
\end{figure}

Due to  these  facts,  we considered the same coupled non-linear oscillators model as in the case of NGC 4151, but with three times smaller  coupling parameter ($cp_{b\rightarrow a}=0.2$).   Fig.~\ref{fig:sim55} depicts  correlation structure of  simulated signals. There is pronounced elongated cluster at prevailing period of 500 days. The  other period of 300 days is more subtle  due to  its association with smaller coupling coefficient.

\begin{figure*}
\centering
\begin{tabular}{cc}
    \includegraphics[width=0.47\linewidth]{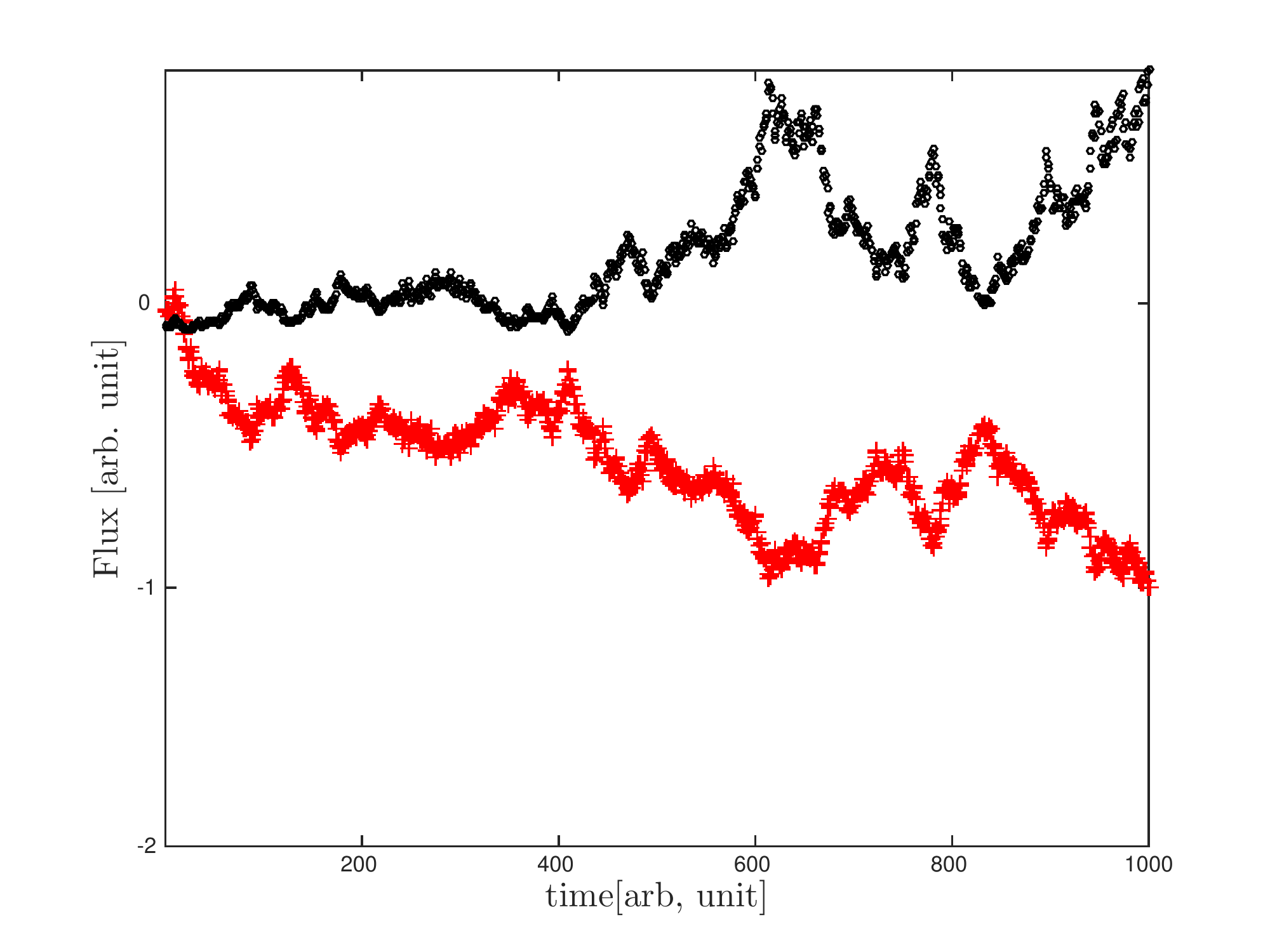}&
    \includegraphics[width=0.47\linewidth]{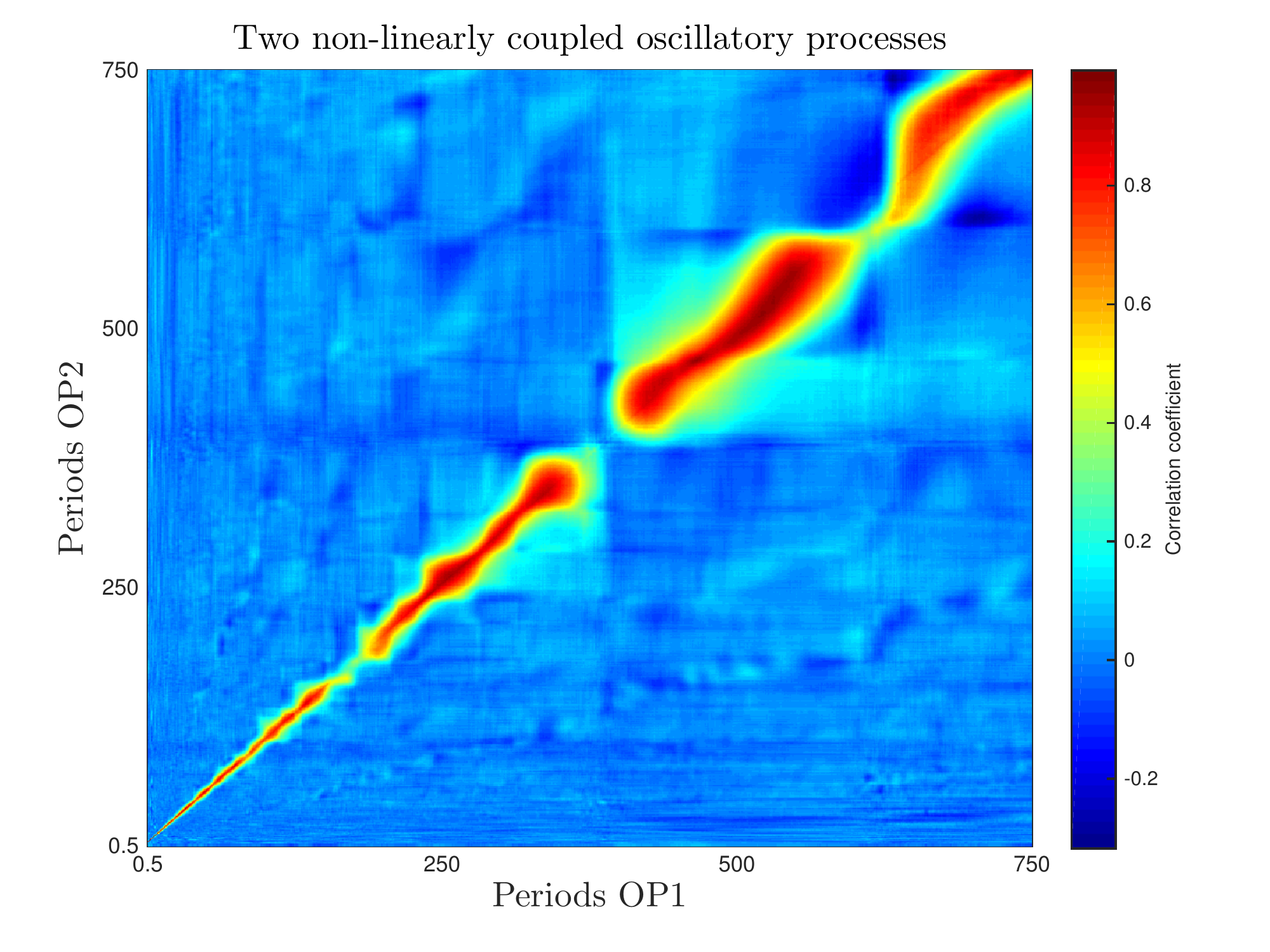}\\  [2\tabcolsep]
 \end{tabular}
\caption{Simulation of   two bidirectional coupled oscillators  for the case of  NGC 5548. Left:  
Random realization  of  Eq.~(\ref{eq:eq8}) 
  form two time series (black is $U_{a}=OP1$, and red is $U_{b}=OP2$) of amplitudes    $A=5.92, B=1.27$, phases $\phi=2.65\, \, \mathrm{ rad}$, coupling strengths $cp _{a\rightarrow b}=0.7, cp _{b\rightarrow a}=0.2$, periods 500, 300  and time delay is 100 arbitrarily chosen time units,. Right:  corresponding  2D correlation map.}
  \label{fig:sim55}
\end{figure*}

Dominant presence of larger period (i.e lower frequency) is similar to 
situation described  in   \citet{Far14} where  was simulated  two-dimensional (2D) hydrodynamical simulations of circumbinary disc accretion  in the binary black hole system.
 They found low frequency mode   as dominant component for the case of  binary black hole  mass ratio $q\gtrsim 0.43$ and that  it  
  corresponds   to the orbital period of a lump in the inner circumbinary accretion disc, rather than  to the orbital binary  period.
However, their simulation revealed that  low frequency  is associated with cadence of  higher  frequency harmonics. Thus, the  scenario of the orbiting lump within the circumbinary disc  is more applicable with regard to the cadence of periodicities found in the case of NGC 4151 (see Fig. ~\ref{fig:per41}).

The predominant periodic oscillation ($\sim$ 12 yr, i.e. $\sim 5000 $ days)    can influence other physical properties of this object, i.e.  the size of the  BLR presumably. 
From the inspection of Table \ref{tab:zdcf}, it is apparent significant variability of the H$\beta$
lags. The time evolution of  ZDCF lags shown in  Fig. ~\ref{fig:blr} resembles almost periodic behavior  of 4600 days,  judging by only the data points.
 However, some of the error bars are very large indicating that within the errors there is no periodicity.
We noticed that the largest error bars occurred for the time windows where the dissimilarity between portion of two light curves is evident and/or a few events (flares or dips) are presented in  one portion but absent in other. The flares are predominant in the continuum within time windows MJD 42500-43250, 43250-44000, 44100-44900, for which the errorbars are the largest.
Possible variations in the H$\beta$ lag have been noticed earlier. Variations of the BLR size of NGC 5548 are between $\sim 5$ and $\sim 30$ light days  and they can arise due to 
numerous physical reasons\citep[see][]{Pet99,Lu16}. 

Moreover, in Fig. ~\ref{fig:ph55} has  been shown  the phase space trajectories of the continuum and simulated curve. 
\begin{figure}
    \includegraphics[width=\columnwidth]{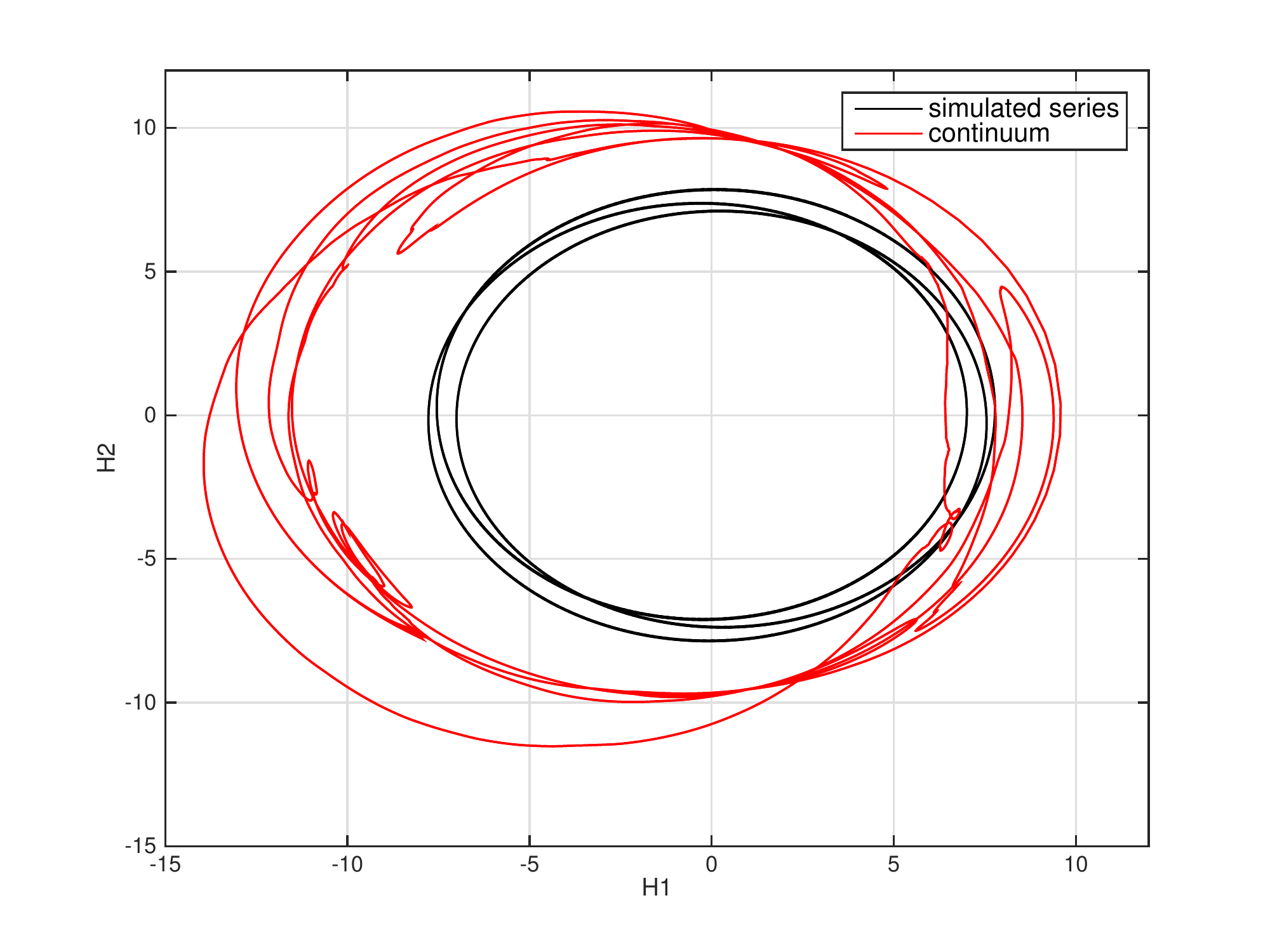}
\caption{As in Fig. ~\ref{fig:ph3c} but for  the continuum  of NGC 5548 and  simulated curve (details are given in the text). Note chaotic like appearance of both curves.} 
   \label{fig:ph55}
\end{figure}
In difference to previous cases,  distinct periodic waveforms  are not present.
Namely,  instead of regular oscillation, 
the presence of several limit cycles forming  a wide ring is observed which is somewhat similar to the chaotic behavior with random wandering of the states in the phase space.
This is in accordance with the large correlation cluster seen in correlation maps of  the real  (Fig. ~\ref{fig:per55}) and simulated curves (Fig. ~\ref{fig:sim55}). Namely, we can suspect that if the amplitude of driving oscillator is strong enough, 
 period overlapping occurs, which is  triggering 
chaotic behavior. This is in line with the fact that chaos can be  born from an overlap of fundamental frequencies of  an unperturbed system \citep{Ter04}.

For  E1821+643, \citet{Kov17}  derived averaged periods  of 4.78, 5.83 and 12.19 yr  from all models  of the continuum 5100 \AA\,, 4200\AA\,,  and  H$\gamma$ emission line. 
However they could not conclude anything about periodicity in the H$\beta$ emission line due to presented large noise. 
Its 2D correlation maps are quite similar to the case of NGC 4151. Particularly, if  we look at  phase portraits of the light curves (Fig. ~\ref{fig:phe})  normal limit cycles are observed in the dynamics of E1821+643.   They are quite similar to phase portrait of regular sinusoids. We note the presence of  two smaller elongated loops in all phase curves  reflecting two smaller periods.
\begin{figure}
    \includegraphics[width=\columnwidth]{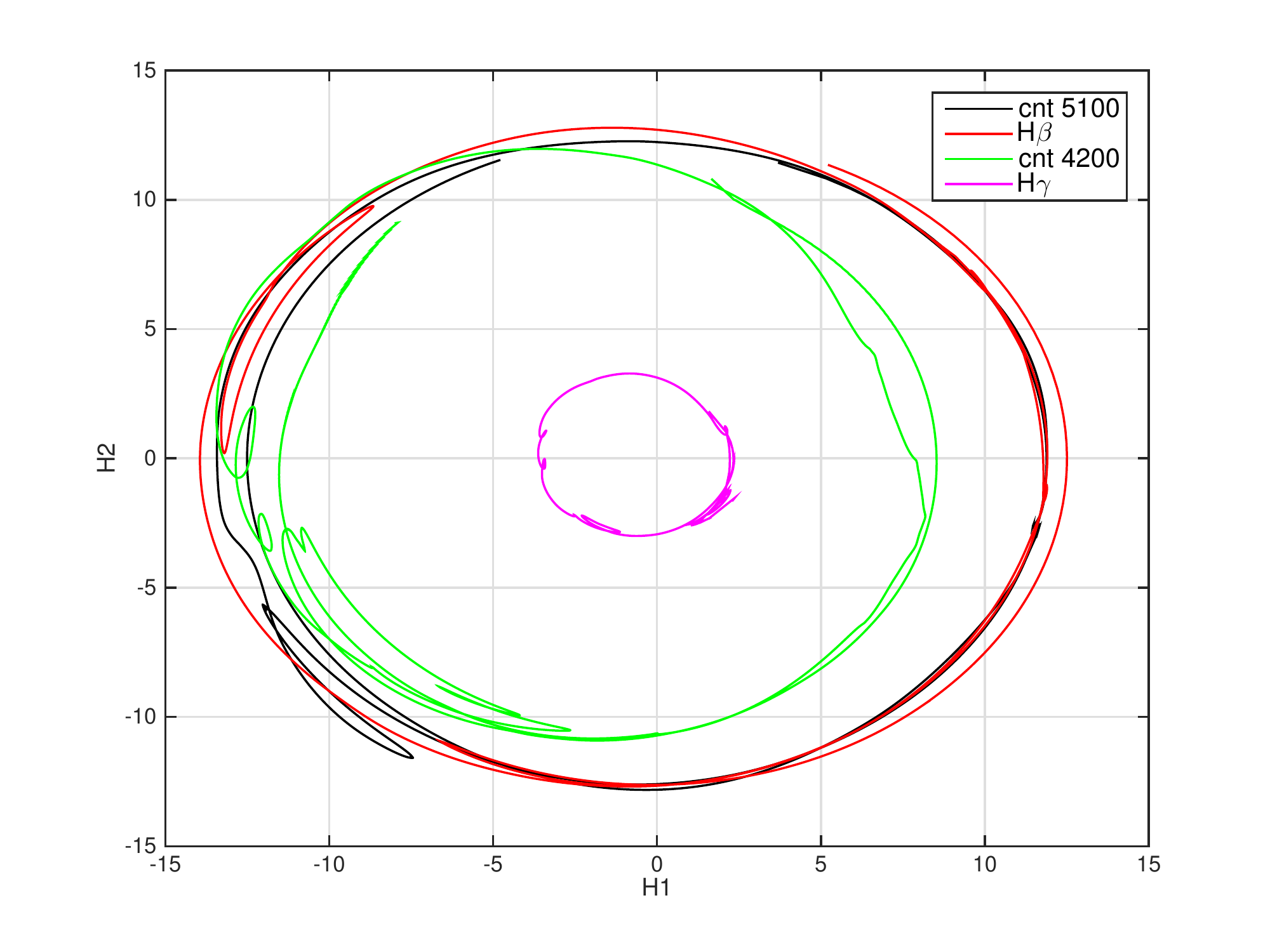}
\caption{As in Fig. ~\ref{fig:ph3c} but for   the light curves   of  E1821+643  (details are given in the text). Topology  of phase curves is  similar to the  phase trajectory of  regular sinusoids. Note that here the width of rings is drastically reduced in comparison with  phase trajectory of NGC 5548 (Fig. ~\ref{fig:per55}), so the system  is more closer to the simple oscillations.} 
   \label{fig:phe}
\end{figure}
Also, boundedness of the phase portraits confirms the stability  of process.
One can think that   small variations in the light curves of E1821+643 \citep[e.g. for the H$\beta$ line F$_{var} \sim 7\%$, see][] {Il17},  can be  a reason 
that  oscillatory patterns are not distorted.
Based on  the above discussion, NGC 5548 and E1821+643 represent dynamical  extremes, the most chaotic and stable in our sample, respectively.

The  correlation of oscillatory patterns  in the  continuum and emission lines  of  all objects (except Arp 102B), can be a consequence of  a more  general correlation trend between
 the continuum and emission lines fluxes  of these objects \citep[see discussion for  correlation between continuum and H$\beta$ emission lines of these objects in][] {Il17}.
In addition, as  time coverage of observations of  NGC 4151 continuum is shorter then time coverage of observations of  the  H$\alpha$ line, 
our study  did  not investigate how the weak (or even absent in some time periods)  correlation between the continuum and emission line fluxes  of this object \citep[see details in][]{Sh08} would  affect the detection of  correlation of  their  oscillatory patterns.
Finally, to investigate the periodic characteristics of the sample, we  correlated logarithmic  values of  obtained periods with  AGN optical luminosities and with their BLR  sizes.
Our findings of absence of any such correlation trend,   do not contradict  current understanding of  relations between those physical parameters of  AGN  \citep{Lu16}, but show that it   is  independent  of 
correlation between continuum and emission line fluxes  if presented in the sample. 
 Moreover, the  fitting of multiple sinusoids to the observed light curves (see Table~\ref{tab:resultfit} and Fig. \ref{fig:sinusfit}) identified   periodicities very similar to those obtained from hybrid method (see Table~\ref{tab:result}). Thus, all fitted multisine models (except of poor fit of NGC 5548 data) unanimously support the existence of periodicities detected by proposed  hybrid method. 
 
It is important to acknowledge that  in addition to  the signal to noise ratio, errors and
irregular sampling,  the stochastic behavior in
red-noise dominated AGN light curves can easily imitate periods with less than several repetitions, so some of  detected periodicities could be due to this effect.
It  does not  imply that the behavior is physically unrealistic, or that it is not 
occurring  in  the lines and the continuum in a causal way, just that it is 
not part of a periodic oscillation. 

Whether or not  all  (or particular)  AGN variability could be caused by superposition of many oscillations is  an open and controversial question and beyond the scope of our work.
 However, we  note the oscillating structures have been modeled \citep[][]{TB08} as particle like (blobs or hot spots, with or without resonant 
interactions) and waves (oscillating tori, discoseismic waves).
Even derived analytic solutions for accretion discs  are  stable against 
finite perturbations, these perturbations could still  excite oscillatory behavior. A 
local restoring  forces within accretions discs can rise oscillations \citep[for a details see review][and 
references therein]{AF13} such as: local pressure gradients running via sound waves, 
 buoyancy forces  operating via   gravity waves, the Coriolis force  acting  through inertial waves, and 
  surface waves appearing  due to the local effective gravity. Such mechanisms
    have been argued to  power the quasi-periodic oscillations. Particularly,  families of  oscillations of 
    low order modes are of special interest  since they could exist in different  accretion disc geometries.
     Such modes will show  tendency  to have the largest amplitudes and produce more visible variations than
      their higher-order complements \citep[][]{AF13}.

\section{Conclusion} 

We searched for oscillatory patterns in the new  combined light curves of 5 well known type 1AGN using a novel hybrid method and oscillatory network models.
The results of our analysis are summarized as follows:
\begin{enumerate}
\item We find strong evidence for periodicities  in the combined light curves of 3C 390.3,  NGC 4151, NGC 5548 and E1821+643.  
The absence of oscillatory patterns in the combined   light curves of Arp 102B distincts clearly this object from other.
 \item We constructed the coupled oscillatory models which  resemble detected oscillatory behavior in the light curves and  confirm their physical background. In the case of Arp 102B the absence of periodicities can be
  concealed   by either  unfavorable time series characteristic or   sufficiently weak coupling   between oscillators mechanisms as our model suggests.
\item We demonstrate that dynamics of two binary black hole candidates NGC 5548 and E1821+643 converge to chaotic  and stability regime, respectively.
\end{enumerate}

\section*{Acknowledgements}
The authors sincerely  thank to the Referee for the constructive comments and recommendations
which definitely  improve the  quality of the paper.  L. \v C. P. is grateful to the Alexander von Humboldt Foundation for 
support. 
This work is supported by  projects (176001)  \textit{Astrophysical
Spectroscopy of Extragalactic Objects}, (176002) \textit{Influence of collisional processes on astrophysical plasma line shapes} of  Ministry of Education, Science and Technological Development
of Serbia,   INTAS (grant N96-
0328) and RFBR (grants N97-02-17625 N00-02-16272, N03-
02-17123, 06-02-16843, N09-02-01136,12-02-00857a, 12-02-
01237a, N15-02-02101).The work was partly supported by the
Erasmus Mundus Master Program, AstroMundus.





\begin{thebibliography}{99}

\bibitem[\protect\citeauthoryear{Abramowic \& Fragile}{2013}]{AF13}
 Abramowic  M.~A.,  Fragile, P.~ C., 2013,  Living Rev. Relativ., 16, 1


\bibitem[\protect\citeauthoryear{Afanasiev et al. }{2015}]{A15}
Afanasiev V. ~L., Shapovalova A. ~I.,  Popovi\'c L. ~\v C.,  Borisov N. ~V., 2015, MNRAS,  448,  3, 2879

\bibitem[\protect\citeauthoryear{Aghabozorgi et al. }{2015}]{Ag15}
Aghabozorgi S.,   Shirkhorshidi A. S.,  Wah T. Y., 2015, Informaton systems, 53, 16

\bibitem[\protect\citeauthoryear{Alexander}{2013}]{Al13}	
Alexander T., 2013, arXiv:1302.1508

\bibitem[\protect\citeauthoryear{Altman et al. }{2000}]{Alt00}
Altman  D. G, Machin D.,  Bryant  T. N., Gardner  M. J., 2000,Statistics with Confidence - Confidence Intervals and Statistical Guidelines  (2nd edition), British Medical Journal


\bibitem[\protect\citeauthoryear{Balanov et al.}{2009}]{Bal09}
Balanov  A., Janson  N., Postnov  D., Sosnovtseva  O., 2009, Synchronization: From simple to complex, Springer-Verlag Berlin Heidelberg

\bibitem[\protect\citeauthoryear{Bentz \& Katz }{2015}]{BK15}
Bentz  M.~ C., Katz S., 2015, PASP,  127, 947, 67

\bibitem[\protect\citeauthoryear{Best \& Roberts}{1975}]{BR75}
Best  D. J.,  Roberts  D. E., 1975, Applied Statistics, 24, 377


\bibitem[\protect\citeauthoryear{Bon et al. }{2012}]{Bon12}
Bon E.,  Jovanovi{\'c} P., Marziani  P.,  Shapovalova, A. I.,  Bon  N., Borka Jovanovi{\'c} V. et al.  2012, ApJ, 759, 2, id.118 

\bibitem[\protect\citeauthoryear{Bon et al. }{2016}]{Bon16}	
Bon E.,  Zucker S., Netzer H., Marziani P.,  Bon N.,  Jovanovi{\'c}  P.,  Shapovalova A. ~I.,  Komossa S.,  Gaskell C. ~M., Popovi{\'c} L.~ {\v C}., et al., 2016, ApJSS,  225,  2,  id. 29

\bibitem[\protect\citeauthoryear{ Br{\o}ns et al.}{2008}]{Bro08}
 Br{\o}ns M.,   Kaper T. ~J.,  Rotstein H. ~G., 2008, AIP Chaos, 18, 015101

\bibitem[\protect\citeauthoryear{Conover et al.}{1999}]{Con99}
Conover W.J, 1999, Practical Nonparametric Statistics (3rd edition), Wiley 


\bibitem[\protect\citeauthoryear{Charisi et al. }{2016}]{Ch16}
Charisi M.,  Bartos I.,  Haiman Z.,  Price-Whelan A.~M.,  Graham M.~J.
 Bellm E.~C, Laher  R.~R.,  M\'{a}rka, S.,  2016, MNRAS,  463, 2145


\bibitem[\protect\citeauthoryear{Dietrich et al. }{1998}]{Diet98}
Dietrich M.,  Peterson B.~M., Albrecht  P., Altmann M.,  Barth  A. ~J.,  Bennie  P. ~J., 
 Bertram  R.,  Bochkarev N.~G. et al.,  1998, ApJSS, 115, 2, 185


\bibitem[\protect\citeauthoryear{Dietrich et al. }{2012}]{Diet12}
Dietrich M.,  Peterson B.~M.,  Grier  C.~J.,  Bentz M.~C., Eastman J., et al., 2012,  ApJ, 757, 1, id. 53 


\bibitem[\protect\citeauthoryear{Ding et al. }{2008}]{Ding08}
 Ding H., Trajcevski G.,  Scheuermann P.,  Wang X., Keogh E., 2008, Proceedings of the VLDB Endowment, Vol. 1,  Issue 2, 1542





\bibitem[\protect\citeauthoryear{Fan et al. }{1997}]{Fan97}
Fan  J.~H.,  Xie G.~Z., Lin R.~G., Qin  Y.~P.,  Li K.~H.,  Zhang X, 1997, A\&AS, 125, 525 



\bibitem[\protect\citeauthoryear{Farris et al.}{2014}]{Far14}	
Farris B. D.,   Duffell P.,  MacFadyen A.~I.,  Haiman Z., 2014, ApJ, 783:134



\bibitem[\protect\citeauthoryear{Grebogi et al.}{1987}]{GOY87}
Grebogi C., Ott  E., Yorke J.~A.,1987, Science, 238, 632.

\bibitem[\protect\citeauthoryear{Grier et al.}{2017}]{Gri17}	
 Grier C. J., Pancoast A., Barth A. J., Fausnaugh M. M., Brewer B. J., Treu T., Peterson B. M., 2017, ApJ,  849,  2,  id. 146, 18p

\bibitem[\protect\citeauthoryear{Grinsted et al.}{2004}]{Grin04}	
 Grinstead, A., Moore, J.~C., Jevrejeva, S., 2004, Non-linear Processes in Geophysics, European Geosciences Union
(EGU),  11 (5/6), 561

\bibitem[\protect\citeauthoryear{Guo et al.}{2017}]{Guo17}	
Guo H.,   Wang J.,  Cai Z., Sun M., 2017, ApJ, 847,2,1 


\bibitem[\protect\citeauthoryear{Hollander et al.}{1999}]{Holl99}
Hollander M, Wolfe D.A., 1999, Nonparametric Statistical Methods (2nd edition), Wiley 
 

\bibitem[\protect\citeauthoryear{H{\"o}nig}{2014}]{Hon14}
H{\"o}nig  S. F., 2014, ApJL,  784,  1, L4

\bibitem[\protect\citeauthoryear{Hawkins}{2002}]{Haw02}	
Hawkins  M. R. S.,  2002, MNRAS, 329, 76 




\bibitem[\protect\citeauthoryear{Ili{\'c} et al.}{2017}]{Il17}
Ili{\'c} D.,  Shapovalova  A. I., Popovi{\'c}  L. {\v C}., Chavushyan V., Kollatschny  W. et al., 2017, Frontiers in Astronomy and Space Sciences, 4,12

\bibitem[\protect\citeauthoryear{Jovanovi{\'c} et al.}{2010}]{Jov10}
Jovanovi{\'c}  P., Popovi{\'c}  L. {\v C}., Stalevski  M.,  Shapovalova  A. I., 2010, ApJ, 718,1,168

\bibitem[\protect\citeauthoryear{Kaspi et al.}{1996}]{Kas96}	
Kaspi  S.,  Maoz  D., Netzer H.,  Peterson  B. ~M.,  Alexander T.,  Barth  A. ~J. et al., 1996, ApJ, 470, 336


 \bibitem[\protect\citeauthoryear{Kawaguchi  \& Mineshigei }{1999}]{KW99}
 Kawaguchi T., Mineshige S., 1999,   
 Active Galactic Nuclei and Related Phenomena, Proceedings of IAU Syposium 194, held 17-21 Aug. 1998, in Yerevan, Armenia. Edited by Y. Terzian, E. Khachikian, and D. Weedman, San Francisco: Astronomical Society of the Pacific, 356
 


\bibitem[\protect\citeauthoryear{Keogh et al.}{2001}]{Ke01}
Keogh  E.,  Chakrabarti K.,  Pazzani  M., Mehrotra S., 2001, Knowledge and Information Systems, 3, 3, 263 





\bibitem[\protect\citeauthoryear{Kova{\v c}evi{\'c} et al.}{2017}]{Kov17}
Kova{\v c}evi{\'c}  A., Popovi{\'c}  L. ~{\v C}., Shapovalova  A. ~I.,  Ili{\'c} D., 2017, Ap\&SS,  362, 2, id.31

\bibitem[\protect\citeauthoryear{Kozlowski.}{2017}]{Koz17}
Kozlowski S., 2017, A\&A,  597, id. A128,1

\bibitem[\protect\citeauthoryear{Kralemann et al.}{2008}]{Kral08}
 Kralemann  B.,  Cimponeriu   L.,  Rosenblum   M.,  Pikovsky   A.,  Mrowka   R., 2008,  Phys. Rev. E 77, 066205 

\bibitem[\protect\citeauthoryear{Kralemann et al.}{2011}]{Kral11}
Kralemann  B., Pikovsky  A.,   Rosenblum  M., 2011,  Chaos  21, 025104

\bibitem[\protect\citeauthoryear{Kudryavtseva  \& Pyatunina}{2006}]{KP06}
Kudryavtseva  N. A.,  Pyatunina T. B., 2006, Astronomy Reports, 50,  1,  1

\bibitem[\protect\citeauthoryear{Kudryavtseva et al.}{2011}]{Kud11}
Kudryavtseva N. A.,  Britzen S.,  Witzel, A. Ros E.,   Karouzos M., 2011, A\&A, 526, A51

\bibitem[\protect\citeauthoryear{Lai \& Ye }{2003}]{LY03}
Lai Y.-C., Ye N., 2003, International Journal of Bifurcation and Chaos, 13,  6, 1383


\bibitem[\protect\citeauthoryear{Lainela et al. }{1999}]{L99}
Lainela M., Takalo  L. O., Sillanp{\"a}{\"a} A., Pursimo  T., Nilsson  K. et al., 1999,  ApJ, 521, 2, 561 



\bibitem[\protect\citeauthoryear{Leighly }{2005}]{Lei05}	
Leighly  K. M., 2005, Astrophysics and Space Science,  300,  1-3,  137

\bibitem[\protect\citeauthoryear{Li et al.}{2016}]{Li16}
Li  Y.-R., Wang  J.-M., Ho  L. C., Lu  K.-X., Qiu  J. et al. 2016, ApJ, 822, 4

\bibitem[\protect\citeauthoryear{Liu et al.}{2016}]{Liu16}
Liu  J., Eracleous M., Halpern J.~ P., 2016, ApJ,  817,  1,  id. 42



\bibitem[\protect\citeauthoryear{Lu et al.}{2016}]{Lu16}		
Lu K.-X., Li Y.-R.,  Bi S.-L.,  Wang J.-M., 2016, MNRAS, 459, 1, p. L124

\bibitem[\protect\citeauthoryear{Mushotzky et al.}{2011}]{Mush11}	
 Mushotzky, R. F., Edelson  R., Baumgartner  W., Gandhi  P., 2011, ApJL, 743, 12 P


\bibitem[\protect\citeauthoryear{Nakao}{2015}]{Nak15}		
Nakao  H.,  2015,  Contemporary Physics. 57, 1,

\bibitem[\protect\citeauthoryear{Nekorkin}{2015}]{Nek15}	
Nekorkin  V. I., 2015, Introduction to non-linear oscillations, Willey-VCH, Verlag GmbH \& Co. KGaA

\bibitem[\protect\citeauthoryear{Netzer}{2013}]{N13}	
Netzer  H., 2013, The Physics and Evolution of Active Galactic Nuclei, Cambridge University Press

\bibitem[\protect\citeauthoryear{O' Brien et al. }{1998}]{OB98}	
O'Brien P. ~T., Dietrich M., Leighly K.,  Alloin D., Clavel J.,  Crenshaw D. ~M.,  Horne K. et al., 1998, ApJ,  509, 1, 163


\bibitem[\protect\citeauthoryear{Pancoast et al.}{2015a}]{Panc15a}
 Pancoast A.,  Brewer  B., Treu T., 2015a, MNRAS,  445,  3, 3055

\bibitem[\protect\citeauthoryear{Pancoast et al.}{2015b}]{Panc15b}
 Pancoast A.,  Brewer  B., Treu T., Park D. et al., 2015b, MNRAS,  445,  3, 3055


\bibitem[\protect\citeauthoryear{Pering et al.}{2014}]{Peri14}
Pering   T. D., Tamburello   G. McGonigle   A.~J.~S.,  Hanna   E.,  Aiuppa  A.,  2014, Computers \& Geosciences,  70, 206

\bibitem[\protect\citeauthoryear{Peterson et  al.}{1999}]{Pet99}
Peterson   B. M., Barth  A. J., Berlind    P.,  Bertram  R., Bischoff   K., Bochkarev N. G. et al.,  1999,  ApJ, 510, 659

\bibitem[\protect\citeauthoryear{Peterson et  al.}{2002}]{Pet02}
Peterson   B. M., Berlind    P.,  Bertram  R., Bischoff   K., Bochkarev N. G. et al.,  2002,  ApJ, 581, 197


\bibitem[\protect\citeauthoryear{Pikovsky, Rosenblum \& Kurths}{2001}]{Pi01}
Pikovsky  A., Rosenblum  M., Kurths  J., 2001, 
Synchroniation: A universal concept in non-linear sciences. Cambridge Non-linear Science Series 12 Eds Chirkov B., Moss, F., Cvitanovi{\'c}  P, Swinney  H. Cambridge University Press



\bibitem[\protect\citeauthoryear{Popovi{\'c} et  al.}{2011}]{Pop11}
Popovi{\'c} L. {\v C}., Shapovalova A. I.,  Ili{\'c} D., Kova{\v c}evi{\'c}  A., Kollatschny  W., Burenkov  A. N., et al., 2011, A\&A,  528, id.A130



\bibitem[\protect\citeauthoryear{Popovi{\'c}}{2012}]{Pop12}
Popovi{\'c}  L. {\v C}, 2012, NewAR,  56, 2-3, 74

\bibitem[\protect\citeauthoryear{Popovi{\'c} et al.}{2014}]{Pop14}
Popovi{\'c} L. {\v C}., Shapovalova A. I.,  Ili{\'c} D., Burenkov  A. N., Kollatschny  W.,  et al., 2014, A\&A, 572, id. A66 

\bibitem[\protect\citeauthoryear{Rees}{1984}]{Re84}
Rees  M. J., 1984, ARA\&A, 22, 471


\bibitem[\protect\citeauthoryear{Risaliti \& Lusso}{2017}]{RiL17}
Risaliti, G., Lusso, E., 2017, Astronomische Nachrichten, 338,  2-3,  329




\bibitem[\protect\citeauthoryear{Sergeev et al. }{2000}]{Ser00}
Sergeev S. ~G., Pronik V. ~I., Sergeeva  E. ~A.,  2000, A\&A,  356, 41


\bibitem[\protect\citeauthoryear{Sergeev et al.}{2007}]{Ser07}	
Sergeev S. ~G., Doroshenko V. ~T.,  Dzyuba S. ~A., Peterson B. ~M., Pogge R. ~W.,  Pronik V. ~I., 2007, ApJ, 668,  2,  708

\bibitem[\protect\citeauthoryear{Sergeev et al. }{2011}]{Ser11}
Sergeev S.~ G., Klimanov S. ~A., Doroshenko V. ~T., Efimov Yu.~ S., Nazarov S. ~V.,  Pronik V. ~I., 2011, MNRAS, 410,3,1877

\bibitem[\protect\citeauthoryear{Sesar et al. }{2007}]{Ses07}
Sesar B., Ivezi{\'c} {\v Z}.,  Lupton R. H., Juri{\'c} M., Gunn J. E.,  Knapp G. R., et al., 2007, AJ,  134,  6,  2236



\bibitem[\protect\citeauthoryear{Simm et al. }{2015}]{Sim15}
 Simm  T., Saglia  R.,   Salvato  M.,  Bender  R.,   Burgett  W.~ S. et al., 2015, A\& A, 584, A106


\bibitem[\protect\citeauthoryear{Shapovalova et al.}{2004}]{Sh04}	
Shapovalova A. ~I.,  Doroshenko V. ~T.,  Bochkarev N. ~G., Burenkov  A. ~N., Carrasco L.,  Chavushyan V. ~H.,  Collin S. et al., 2004, A\&A, 422, 925

\bibitem[\protect\citeauthoryear{Shapovalova et al.}{2008}]{Sh08}
Shapovalova  A.~ I., Popovi{\'c} L. {\v C}.,  Collin S.,  Burenkov A. ~N.,  Chavushyan  V.~ H., Bochkarev  N. G. et al., 2008, A\&A, 486,  1, 99 

\bibitem[\protect\citeauthoryear{Shapovalova et al. }{2010a}]{Sh10a}
Shapovalova A.~ I., Popovi{\'c}  L. ~{\v C}.,  Burenkov A. ~N.,  Chavushyan V. ~H., Ili{\'c} D., Kollatschny W., Kova\v cevi{\'c}  A. et al., 2010a, A\&A, 517, id. A42	


\bibitem[\protect\citeauthoryear{Shapovalova et al. }{2010b}]{Sh10b}
Shapovalova A.~ I., Popovi\'c  L. ~{\v C}.,  Burenkov A. ~N.,  Chavushyan V. ~H., Ili{\'c} D., Kova\v cevi{\'c}  A. et al., 2010b, A\&A, 509, id. A106	

\bibitem[\protect\citeauthoryear{Shapovalova et al. }{2013}]{Sh13}
Shapovalova  A. ~I., Popovi\'c L.~ \v C., Burenkov  A. ~N., Chavushyan V. ~H., Ili\'c  D.,  Kollatschny W., Kova\v cevi\'c A.,  Bochkarev  N. ~G. et al., 2013, A\&A, 559, A10


\bibitem[\protect\citeauthoryear{Shapovalova et al. }{2016}]{Sh16}
Shapovalova A.~ I., Popovi\'c  L. ~{\v C}., Chavushyan V. ~H.,    Burenkov A. ~N.,   Ili{\'c} D., Kollatschny W., Kova\v cevi{\'c}  A. et al., 2016, ApJSS, 222, 2, id. 25	

\bibitem[\protect\citeauthoryear{Tassev \&  Bertschinger }{2004}]{TB08}
Tassev S. V.,  Bertschinger E., 2008, ApJ, 686,  1,  423


\bibitem[\protect\citeauthoryear{Tenenbaum et al. }{2000}]{Ten00}
Tenenbaum   J. B.,  De Silva   V., Langford   J. C., 2000, Science, 290,  2319


\bibitem[\protect\citeauthoryear{Terzi{\'c} \& Kandrup}{2004}]{Ter04}
Terzi{\'c} B.,  Kandrup  H. E., 2004,  347, 3,  957


\bibitem[\protect\citeauthoryear{Torricelli-Campioni et al. }{2000}]{TC00}
Torricelli-Ciamponi  G., Foellmi  C., Courvoisier  T. J.-L., Paltani  S., 2000, A\&A, 358, 57



\bibitem[\protect\citeauthoryear{Wamsteker  et al. }{1997}]{W97}
Wamsteker W.,  Wang T.-G.,  Schartel N.,  Vio R., 1997, MNRAS, 288, 225

\bibitem[\protect\citeauthoryear{Wang \& Megalooikonomou}{2008}]{WQ08}
Wang Q.,   Megalooikonomou  V., 2008, Information Systems,  33,  1, 115

\bibitem[\protect\citeauthoryear{Webb  et al. }{1988}]{Webb88}
Webb  J. R., Smith  A. G.,  Leacock  R. J., Fitzgibbons G. L., Gombola, P. G., 1988, In: Supermassive black holes,  Proceedings of the Third George Mason Astrophysics Workshop, edited by M. Kafatos,  Cambridge University Press, 66



\bibitem[\protect\citeauthoryear{Yang \& Tse }{2005}]{YT05}
Yang W, Tse P. W., 2005,  Journal of  Vibration and  Acoustics, 127, 3, 280

\bibitem[\protect\citeauthoryear{Zheng et al.}{2016}]{Zh16}
Zheng  Z.-Y., Butler N. R.,  Shen Y., Jiang L., Wang J.-X.,  Chen X., Cuadra, J., 2016,  ApJ, 827,  1,  id. 56





\end{thebibliography}


\bsp	
\label{lastpage}
\end{document}